\documentclass[%
reprint,
superscriptaddress,
amsmath,amssymb,
aps,
]{revtex4-2}

\usepackage{graphicx}
\usepackage{dcolumn}
\usepackage{bm}

\usepackage[caption = false]{subfig}
\usepackage{xcolor}
\usepackage{float}
\begin{document}

\preprint{APS/123-QED}

\title{Impact of the RNA Binding Domain Localization of the Protein Shell on Virus Particle Stability }

\author{Mohammadamin Safdari}
\affiliation{%
Department of Physics, University of California, Riverside, California 92521, USA
}

\author{Siyu Li}
\affiliation{%
Department of Physics, University of California, Riverside, California 92521, USA
}

\author{Sanaz Panahandeh}
\affiliation{Department of Computational Biomedicine, Cedars-Sinai Medical Center 700 N, San Vicente Blvd., West Hollywood, CA, USA
}

\author{Paul van der Schoot}

\affiliation{Department of Applied Physics and Science Education, Eindhoven University of Technology, Postbus 513, 5600 MB Eindhoven, The Netherlands
}
\author{Roya Zandi}

\affiliation{
Department of Physics, University of California, Riverside, California 92521, USA
}

\date{\today}

\begin{abstract}
The encapsulation of polyanions, whether single-stranded RNAs or synthetic polymers, is primarily driven by attractive electrostatic interactions between the positively charged, structurally disordered RNA-binding domains of virus coat proteins and the negatively charged polyanions. Theoretically, this interaction is often modeled by coarse-graining the charge distribution of the binding domains, either by projecting the charges onto the inner surface of the protein shell or by spreading them across a region representing the capsid lumen where the binding domains are located. In practice, however, the positive charges are not uniformly distributed across the binding domains, which themselves are positioned at discrete, specific sites on the shell surface. Here, we use molecular dynamics simulations to investigate the impact of localized interactions on the most probable or optimal length of the encapsulated polymer, revealing that the specific location of charges along the binding domains plays a significant role, consistent with experimental observations. Comparing the simulations with predictions from a simple mean-field theory taken from the literature, we find that while the general trends are reasonably well captured, quantitative discrepancies arise between the two approaches.
\end{abstract}

\maketitle

\section{\label{sec:level1}Introduction}
Simple icosahedral viruses have been reconstituted {\it in vitro} from their constituent components: coat proteins and a single-stranded genome, which is typically RNA rather than DNA~\cite{comas2012vitro,vaupotivc2023viral,hagan2016recent,ning2016vitro,beren2017effect}. Virus-like particles can spontaneously assemble not only around their native RNAs but also around heterologous RNAs, synthetic polyanions, and even negatively charged colloidal particles ~\cite{perlmutter2015role,stockley2013packaging,panahandeh2022virus}. The main driving force is understood to stem from electrostatic interactions between the negative charges on the cargo and the positive charges on structurally disordered RNA-binding domains that are part of the coat proteins~\cite{cadena2011exploiting}. These RNA-binding domains are often referred to as arginine-rich motifs (ARMs) and are typically (but not always) located at the N-terminal end of the coat protein. These N-terminal domains are positioned on the inner surface of the proteinaceous shell, near the vertices of the icosahedral protein shell~\cite{sun2007core,panahandeh2020virus,zandi2020virus,johnson2000structures}.  

There is evidence that the binding of the genome to the N-terminals induces a partial $\alpha$-helical structure in an otherwise unstructured motif, at least in the case of the Cowpea Chlorotic Mottle Virus (CCMV)~\cite{cadene2014vacuum,battesti2008bmv,van1991conformation}. Conversely, binding may alter the secondary (base-pairing) structure of the ssRNA it encapsulates. Near the inner surface of the capsid, the RNA adopts an icosahedral-like conformation, which is disrupted deeper within the lumen~\cite{nap2014role,hu2008packaging,sun2010genome}. In contrast, the structure of encapsulated synthetic polyanions appears to be much more disordered compared to that of ssRNAs~\cite{van2013impact,kivenson2010mechanisms}.

Many ssRNA viruses package genomes that contain more negative charges than are required to neutralize the charge on the N-terminals~\cite{stockley2013packaging,garmann2014role,erdemci2014rna,erdemci2016effects,french2010long,bovzivc2017varieties}. The same holds true for the encapsulation of synthetic polyanions~\cite{douglas1998host,maassen2019elucidating,wang2006combination,chang2008curvature,Hsiang-Ku}. There are numerous theoretical and simulation studies that aim to explain why such overcharging occurs~\cite{perlmutter2013viral,erdemci2014rna,erdemci2016effects,perlmutter2015mechanisms,kundagrami2008theory,muthukumar201750th,ting2011thermodynamic,hu2008electrostatic,muhren2023electrostatic,bruinsma2016equilibrium}. Indeed, the issue of overcharging is well-known in the context of polyelectrolyte adsorption, and, unsurprisingly, it is frequently approached from a polymer physics perspective~\cite{joanny1999polyelectrolyte,van1984lattice,szilagyi2014polyelectrolyte}. The extent of overcharging, if it occurs, depends on various factors, including the architecture of the polymer chains, their flexibility, and the specific approximations used in the models. The accuracy of these predictions remains highly debated, especially when based on the Poisson-Boltzmann theory~\cite{fingerhut2021contact,kirmizialtin2012ionic,nguyen2017accurate,netz2001electrostatistics}.

Even in the context of encapsulating linear polyanions, which are structurally simpler than ssRNAs, current approaches often rely on coarse-graining methods that neglect the physical presence of the N-terminals. These models typically represent the N-terminals as smeared-out charges on a smooth, spherical surface or as a shell near the inner wall of the spherical capsid, overlooking the actual structure of the extended N-terminals of the capsids~\cite{kronenberg2001electron}. A few studies have explicitly modeled the N-terminals in one form or another~\cite{van2013impact,li2017impact,forrey2009electrostatics,wu2003biological,adachi2019n,wu1995structural,ni2012examination,ting2011thermodynamic}. Using field-theory techniques and a more realistic representation of capsids, Li {\it et al.} \cite{li2017impact} and Dong {\it et al.} \cite{dong2020effect} showed that the localization of charges significantly impacts the optimal length of encapsulated polyanions, thereby explaining why overcharging by linear polyanions can indeed be observed experimentally~\cite{schwer2004targeting,tresset2009multiple,knobler2009physical}.

A comprehensive investigation is still lacking regarding the impact of the size of the capsid lumen, the length of the N-terminal regions, the strength of interactions between the N termini and polymeric cargo, and the distribution of charges along the N-terminal backbone. Moreover, the number of amino acids constituting the N-terminal, as well as the number and distribution of charged amino acids, varies between viruses. It is noteworthy that the size of the lumen varies not only across viruses with different T numbers but also among those with the same T number~\cite{lovsdorfer2013statistical,bovzivc2017varieties,lovsdorfer2018anomalous,bovzivc2018compactness}. There appears to be a roughly linear relationship between the number of charges per RNA binding domain and the length of the genome, at least for genomes up to six thousand nucleotides~\cite{belyi2006electrostatic,ting2011thermodynamic,lovsdorfer2012simple,bovzivc2017varieties,lovsdorfer2018anomalous}. This is consistent with the findings of Ni {\it et al.~}\cite{ni2012examination}, who increased the number of positive charges on the RNA binding domains of the simple plant virus Bromovirus (BMV) and observed a corresponding change in the length of the encapsulated genome, despite considerable scatter in the data. Their work shows that the total length of encapsulated RNA is influenced not only by the number of positive charges on the N-terminal domains but also by additional factors, such as the length of the RNA-binding domain and the specific positioning of the positive charges within that domain~\cite{ni2012examination}. 

From a theoretical perspective, nearly all prior attention has been directed toward understanding the impact of the properties of genomes~\cite{ni2012examination,li2017impact,sivanandam2016functional,van2005electrostatics,van2013impact,erdemci2016role}, with much less focus on the properties of the RNA binding domains. The primary goal of this paper is to address this gap, at least in part, by presenting a computational simulation model aimed at investigating the relationship between the amount of encapsulated RNA and the number and arrangement of positive charges on the N-terminal binding domains. In our coarse-grained model, the N-terminal regions are represented as short polymeric chains that, in addition to excluded volume interactions, exhibit attractive interactions with the genome. We vary the distribution of positive charges along the chain to explore their impact on the length of the encapsidated genome. For simplicity, we consolidate all interactions into effective Lennard-Jones-type potentials rather than explicitly addressing the electrostatics of the system, in order to concentrate on the influence of charge distribution along the N-terminal while avoiding the complexities associated with accurately modeling Coulomb interactions, which are notoriously challenging~\cite{muhren2023electrostatic,maassen2018experimental,moghaddam2019hofmeister,cacace1997hofmeister}.

The remainder of this paper is organized as follows. Section II describes the simulation methods employed in our investigations, detailing the computational framework and key assumptions. Section III presents the results obtained from our simulations, accompanied by a thorough analysis and interpretation of the findings. Finally, Section IV highlights the significant trends and patterns identified in the study and discusses their broader implications in the context of our research objectives.

\section{Method and Simulation}
\begin{figure}[h]
	\begin{center}\includegraphics[width=6.cm]{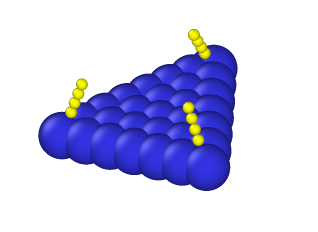}
		\caption{Image of a subunit of our model capsids. The blue particles represent wall particles, which enforce excluded volume effects (ranging from 21 to 55 particles per subunit). The yellow particles represent the N-terminal binding domains that attract the genome (ranging from 1 to 8 particles per N-terminal).}	
		\centering 
		\label{subunits}
	\end{center}
\end{figure}

In this section, we outline the methods and algorithms used in our molecular dynamics (MD) simulations. The aim of this study is to determine the optimal genome length encapsulated by the capsid as a function of the N-terminal domain's length and charge distribution, capsid size, and $T$ number structures. Our analysis includes both $T=1$ and $T=3$ icosahedral capsids, which are composed of 20 and 60 triangular subunits, respectively~\cite{panahandeh2018equilibrium}. Each triangular subunit consists of beads (blue particles shown in Fig.~\ref{subunits}) with a radius of 2.0 nm. The distance between the centers of adjacent beads is 1.0 nm, resulting in overlapping configurations that prevent any material inside the capsid from escaping. The number of beads in each triangular subunit depends on the subunit's size, which can vary by altering the $T$ number or changing the capsid radius for a fixed $T$ number.

For the smallest $T=1$ capsid, each subunit contains 21 beads, while the largest $T=1$ capsid has 31 beads per subunit. Similarly, in $T=3$ capsids, the smaller structures have 31 beads per subunit, whereas the larger ones contain 55 beads per subunit.

We model the genome as a linear bead-spring chain with a variable length ranging from $100$ to $2000$ beads. Each bead in the chain has a radius of $0.5$ nm, which approximates the Kuhn length of single-stranded RNA and the average length of hybridized sections~\cite{yoffe2008predicting,chen2012ionic}. The interaction between two neighboring beads, labeled 1 and 2, along the chain is represented by a harmonic spring potential:

\begin{equation}
U_\mathrm{s} (\vec{r}_1-\vec{r}2) = \dfrac{1}{2} k_\mathrm{s} \left(|\vec{r}_1-\vec{r}_2| - l\right)^2,
\end{equation}
where the energy is expressed in units of thermal energy, $k_\mathrm{B}T$ with $k_\mathrm{B}$ the Boltzmann constant and $T$ the absolute temperature. In this equation, $\vec{r}_1$ and $\vec{r}_{2}$ represent the positions of the two neighboring beads. The spring constant $k\mathrm{s}$ is set to $1000 \ \mathrm{nm}^{-2}$, and $l$, the equilibrium length of the bond, is fixed at 1 nm.

We model the RNA-binding domain (the N-terminal) as a short, stiff chain composed of a variable number of beads, each with a radius of $0.5$ nm, ranging from 1 to 8 beads per N-terminal. These chains are oriented toward the center of the capsid and attached to the vertices of the capsid wall, mimicking the structure of brome viruses (see the yellow beads in Fig.\ref{subunits}). For a $T=1$ capsid, there are 12 vertices, while for a $T=3$ capsid, there are 32 vertices\cite{smith2000structure,silva1985refinement,choi1997structure,fisher1993ordered,tong1993refined,speir1995structures}.

The interaction between a genomic bead and an RNA-binding domain bead, expressed in units of thermal energy, is given by
\begin{equation}
U_\mathrm{LJ} (\vec{r}_1-\vec{r}_2) =
4\varepsilon_{12}\left[\left(\dfrac{\sigma_{12}}{|\vec{r}_1-\vec{r}_2|}\right)^{12} - \left(\dfrac{\sigma_{12}}{|\vec{r}_1-\vec{r}_2|}\right)^6\right],
\label{LJ}
\end{equation}
if $r < r_\mathrm{c}$, and $U_\mathrm{LJ}=0$ for $r \ge r_\mathrm{c}$, where $r_\mathrm{c}$ is the cut-off distance. The parameter $\sigma_{12} = 2^{-1/6}(r_1 + r_2)$ is proportional to the sum of the radii $r_1$ and $r_2$ of the interacting particles.  The parameter $\varepsilon$ controls the strength of the interaction between the beads, and its value is varied from $0.1$ to $1$. For interactions between a genomic bead and an N-terminal bead, we set $r_\mathrm{c} = 10$. For interactions involving non-neighboring genomic beads or between a genomic bead and a wall bead, we use a truncated Lennard-Jones (LJ) potential, retaining only the repulsive component by setting $r_\mathrm{c} = r_1 + r_2$. In these cases, the interaction strength $\varepsilon_{12}$ is fixed at $0.01$. A detailed summary of these parameters is provided in Table~\ref{parameters_LJ}.  Unless specified otherwise, all N-terminal beads interact attractively with the chain beads. In a subsequent section, we relax this assumption by restricting the attraction to only a subset of the N-terminal beads. This modification allows us to explore how localized interactions influence the optimal length of the encapsulated chain.

To determine the optimal genome length for a given capsid, we performed molecular dynamics (MD) simulations using a long polymer chain and a preassembled capsid, where one subunit was made permeable to the polymer, allowing it to either exit or remain inside the capsid. The simulation box measures $200$ $nm^3$ with periodic boundary conditions applied. Instead of focusing on the kinetics of shell assembly \cite{Willy2022,Timmermans}, we concentrate on determining the optimal genome length for encapsidation by a rigid preformed capsid.  During simulations, only the genome particles were allowed to dynamically evolve and relax, while the capsid was kept rigid, without any fluctuations in shape or size. This setup enables the polymer to freely equilibrate, translocate, or exit the capsid. Initially, the polymer was placed inside the capsid in a helical configuration around a hypothetical sphere to facilitate faster relaxation.

Each simulation was run for approximately 50 million steps, during which we monitored the number of genome beads encapsulated within the capsid to identify the optimal genome length. If the entire genome remained inside the capsid during a simulation, we incrementally increased the genome length and repeated the process. This procedure continued until we found a chain length that partially or completely escaped the capsid. The \textit{optimal genome length} is defined as the longest genome that remains fully encapsulated during the simulation.

Through these MD simulations, we identified the thermodynamic optimum polymer length by varying the genome size. Hagan and collaborators previously validated this approach, demonstrating that it reliably predicts the optimal genome length when compared to alternative methods. Incorporating the interaction models described earlier and employing the Langevin integrator allowed us to study how factors such as N-terminal domains, genome interactions, wall particles, and capsid vertices influence the optimal genome length.

The simulations were conducted on an NVIDIA GeForce RTX 3090 GPU. For the MD simulations, we utilized the HOOMD-blue toolkit, a powerful platform for particle-based simulations~\cite{anderson2020hoomd}. Visualization of simulation outcomes was performed using OVITO software~\cite{stukowski2009visualization}.

\begin{figure}[h]
	\begin{center}\includegraphics[width=8cm]{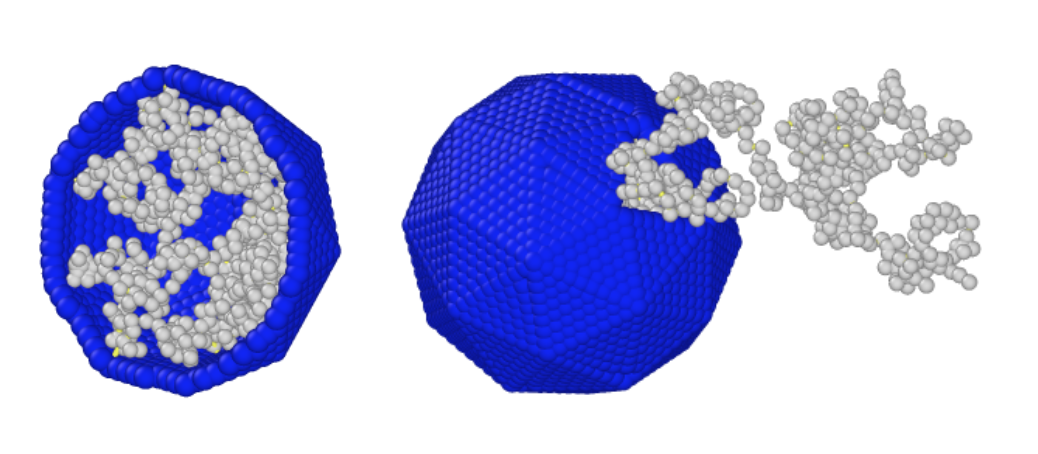}
		\caption{Snapshot of simulations showing a $T=3$ capsid, where part of the genome has escaped through one of the triangular subunits that has been made transparent to the genome. The genome particles are depicted in gray, while the blue particles represent wall particles interacting with the genome through excluded volume effects. The genome is wrapped around yellow particles (representing N-terminal domains), which are not visible but attract the genome. (a) A cutaway view of the capsid, showing the genome inside. (b) The full capsid, with the genome translocating through a missing triangular subunit. }	
		\centering 
		\label{fig0}
	\end{center}
\end{figure}

We find that the genome escapes the capsid entirely when it is either too long or when the attractive energy $\varepsilon$ between the N-terminal and chain beads is too small. A snapshot from a simulation run, presented in Fig.~\ref{fig0}, illustrates that the genome can escape from the shell when it is too long. A large genome may partially escape, with some portion remaining outside the capsid, while an excessively large genome escapes completely.

\begin{figure}[h]
	\begin{center}
		\includegraphics[width=8cm]{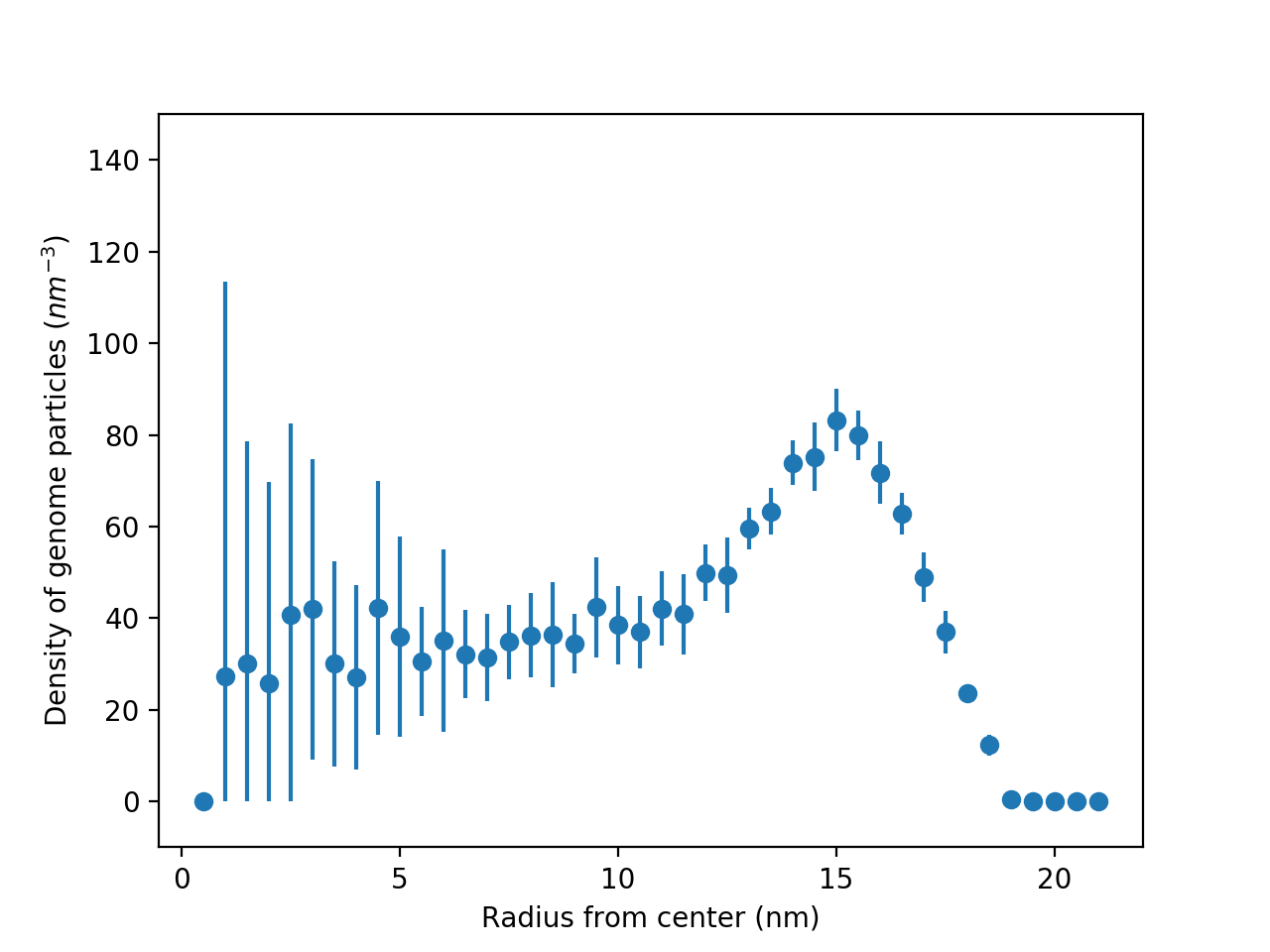}
		\caption{Density of the genomic beads (in units of nm$^{-3}$) as a function of the radial distance from the center of the capsid for a $T=3$ capsid and inner radius $R=21$ nm. The N-terminals consist of six beads. The increment in radius is $0.5$ nm. The interaction strength between the N-terminal beads and the chain beads is $\varepsilon =  0.20$. Indicated are also the standard deviations. }	
		\centering 
		\label{Density_radius}
	\end{center}
\end{figure}

For cases where the chain remains encapsulated, its segments predominantly accumulate near the surface of the capsid, where the N-terminals are located. By plotting the density of genomic beads as a function of the radial position, Fig.~\ref{Density_radius} reveals that most of the genome is wrapped around the N-terminals. To obtain Fig.~\ref{Density_radius}, we count the number of beads in consecutive radial shells, divided by their respective volumes, using a shell width of 0.5 nm. This process begins after 50 million simulation steps, with measurements taken every one million steps. After 10 million steps, we calculate the average density and standard deviation.

The resulting density distribution peaks at the positions corresponding to the ends of the N-terminals and gradually decays to a nearly constant value as it approaches the center of the cavity.

\begin{table}[]
\centering
\begin{tabular}{|c|c|c|c|c|c|}
\cline{1-6}
Particles&$ \ r_1$ \ & $\ r_2 \ $ &$\ r_c \ $& $\sigma_{12}$ & $\varepsilon_{12}$
\\
\hline
g - g&$\ 0.5 \ $& $ \ 0.5 \ $ & $1.0$&$\ 2^{-1/6}\cdot(1.0) \ $&$ \ 0.01 \ $
\\
\hline
g - n&$0.5$& $0.5$ & $10$&$2^{-1/6}\cdot(1.0)$&$0.1\leq \varepsilon \leq 1.0$ 
\\
\hline
g - w&$0.5$& $2.0$ & $2.5$&$2^{-1/6}\cdot(2.5)$&0.01
\\
\hline
\end{tabular}
\caption{Parameter values for the Lennard-Jones interaction between every two particles; g refers to a genome particle, n to an N-terminal particle, and w to a wall particle. Lengths $r_1$, $r_2$ and $r_c$ as well as $\sigma_{12}$ are in nm, and energies $\varepsilon_{12}$ in units of thermal energy. The interaction between the genome and the N-terminal varies between 0.1 and 1.0.}
\label{parameters_LJ}
\end{table}

In the following section, we summarize the main results of our simulations.

\section{Results}
Before presenting our simulation results, we find it instructive to see what a relatively simple mean-field Flory theory predicts for the optimal chain length, $M$, in units of numbers of beads, using in essence a highly simplified and smeared-out version of the model used in our simulations~\cite{van2013impact}. Let $n$ be the total number of charged amino acids per RNA binding domain, of which there are $N$, and let $R$ be the radius of the lumen of the capsid. Then, if we model the N-terminals as short polymer chains consisting of $n$ segments, the Flory theory of~\cite{van2013impact} predicts we must have $M \sim \varepsilon \hspace{0.05cm} n N $ \textit{independent} of the capsid radius $R$\footnote{The model of ~\cite{van2013impact} maps onto the present model by $\varepsilon \rightarrow - \nu_\pm / l^3$ where $\nu_\pm <0$ is the cross virial coefficient between an N-terminal segment and a chain segment and $l$ denotes the bond length.} A similar relation has been found in a density functional theory~\cite{van2005electrostatics}, showing the linear relationship between the optimal length of the encapsulated chain and the total number of charges on the N-terminals.
Unfortunately, this expression is only valid if the attraction between an N-terminal bead and a genome bead is sufficiently weak and $\varepsilon < 1/n$, implying that its range of applicability is somewhat limited. An explicit analytical result for larger adsorption strengths is not available at this time.

If we ignore this limitation for the sake of argument, we can conclude the following from this simple analysis.
\begin{itemize}
    \item[i)] If the interaction between a single bead in the chain and the entire N-terminal is kept constant, such that the product $n\varepsilon$ remains unchanged, the optimal degree of polymerization does not increase with the length of the N-terminals.
    \item[ii)] Increasing the radius of the capsid while keeping all other parameters constant should not lead to an increase in the optimal encapsulated length of the polymer.
    \item[iii)] The number of encapsulated polymer segments increases linearly with the attractive interaction energy $\varepsilon$.
    \item[iv)] The optimal length increases linearly with the T number if the number of N-terminals remain the same. 
\end{itemize}
As we shall see, while these trends are by and large captured by the theory, the predicted scaling exponents are quite inaccurate. This is in part because we venture outside of the range of validity of the theory, but, more importantly, because of the highly course-grained and mean-field nature of the theory. We shall investigate all of these predictions point-by-point next.

\subsection{Effect of the length of the N-terminals}

\begin{figure}[h]
	\begin{center}
		\includegraphics[width=8cm]{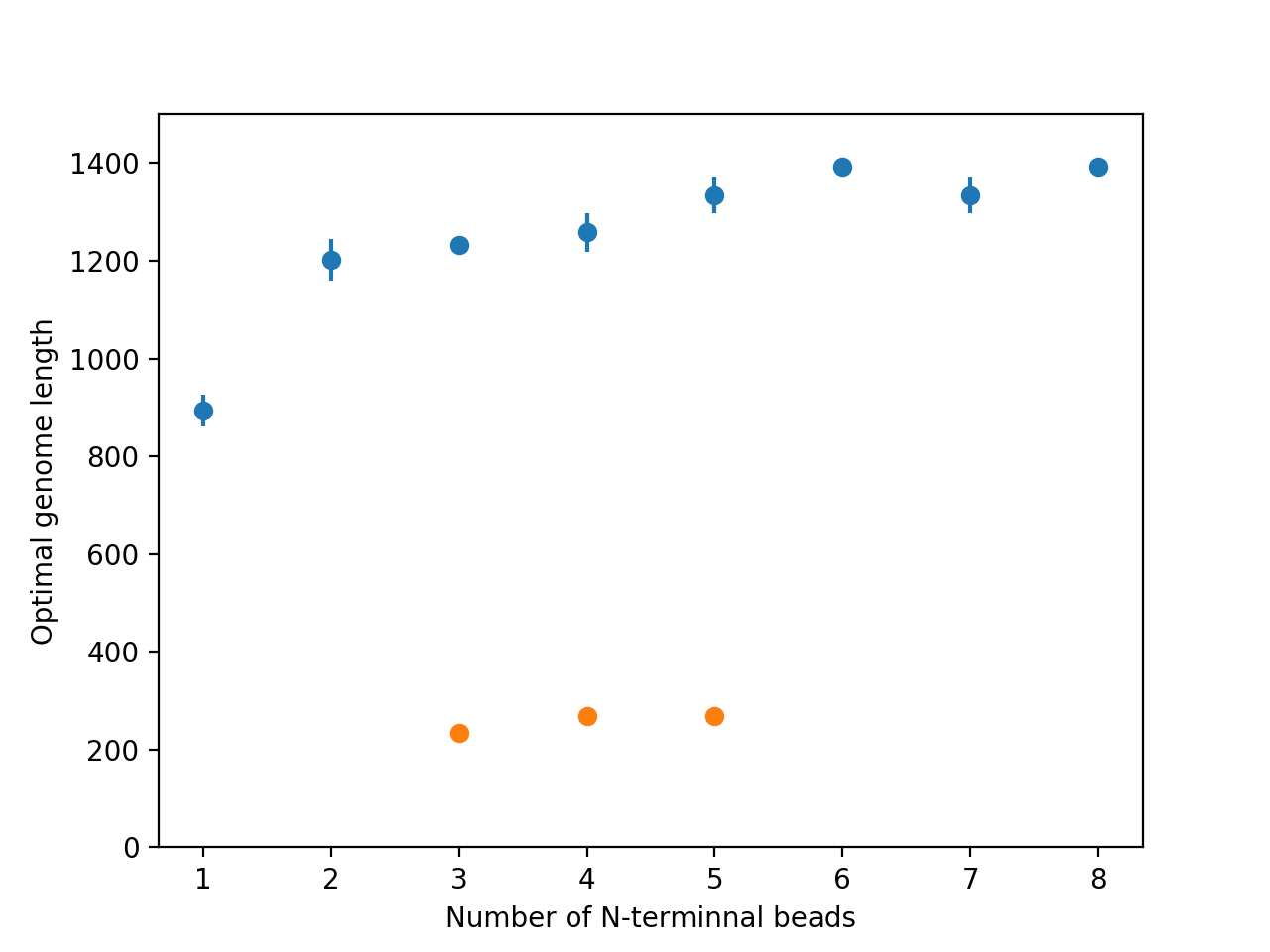}
		\caption{Optimal length of the encapsulated genome as a function of the number of beads $n$ of the N-terminal for a $T=3$-type (blue curve) capsid of inner radius $R=21$ nm and $T=1$-type (orange curve) capsid of inner radius $R=12$ nm. The overall effective interaction energy between a chain bead and an N-terminal is fixed at $n \varepsilon = 1.24$. See also the main text. For the $T=1$ shell, no encapsulation takes place for $n\leq 2$ for the given interaction energy, whilst for $n\geq 6$ the N-terminals fill up the cavity preventing the insertion of the genome.} 
		\centering 
		\label{N-Terminals_length}
	\end{center}
\end{figure}

Figure~\ref{N-Terminals_length} shows that for both $T=3$ and $T=1$ capsids, the optimal encapsulated chain length, $M$, increases with the length of the N-terminals, $n$, even when the total effective binding energy, $\varepsilon n$, is kept constant. For this figure, we set $\varepsilon n = 1.24$ arbitrarily and kept the capsid radius and the number of N-terminals fixed: $R = 21$ nm and $N = 32$ for the $T=3$ capsid, and $R = 12$ nm and $N = 12$ for the $T=1$ capsid. These radii are slightly larger than those of a typical $T=1$ or $T=3$ capsid. We will later examine the effect of the capsid radius on the optimal encapsulated chain length and find that the trend remains the same. If we fit a scaling relation between the optimal chain length and the number of N-terminal beads for the $T=3$ capsid, admittedly over less than a decade in $n$, we obtain a scaling exponent of approximately one-fifth. This is not equal to the value of zero we find from the theory, but does underline the weak dependence on the length of the N-terminals. This (slight) discrepancy is not all that surprising, considering that the effect of localization by adsorption of the chain is very different for the course-grained Flory theory and our simulation model. Polymer adsorption is very much affected by the geometry of the adsorbing surface~\cite{Piculell1995}. It is perhaps not surprising that strong adsorption onto a single bead is less effective at capturing a large polymer than a string of beads with proportionally weaker interactions, as stronger localization leads to a greater reduction in configurational entropy, as shown by our simulations.

For the smaller $T=1$ capsid, we observed no encapsulation for N-terminals consisting of fewer than three beads. For N-terminals with three or more beads, there is a weak dependence on the length of the N-terminals, but the optimal length is much smaller, which is expected given the smaller radius. According to theory, we would anticipate that the optimal encapsulated chain length is smaller by a factor of 3. 
However, we find that it is smaller by a factor closer to 4. This difference may be due to packing constraints, which are more significant for the smaller shell. These packing constraints actually prevent us from investigating N-terminals larger than five beads, as there is simply not enough room to insert a chain. This may also explain why the linear dependence of the optimal chain length does not scale exactly with the $T$ number.

\subsection{Effect of interaction strength}

\begin{figure}[h]
	\begin{center}
\includegraphics[width=8cm]{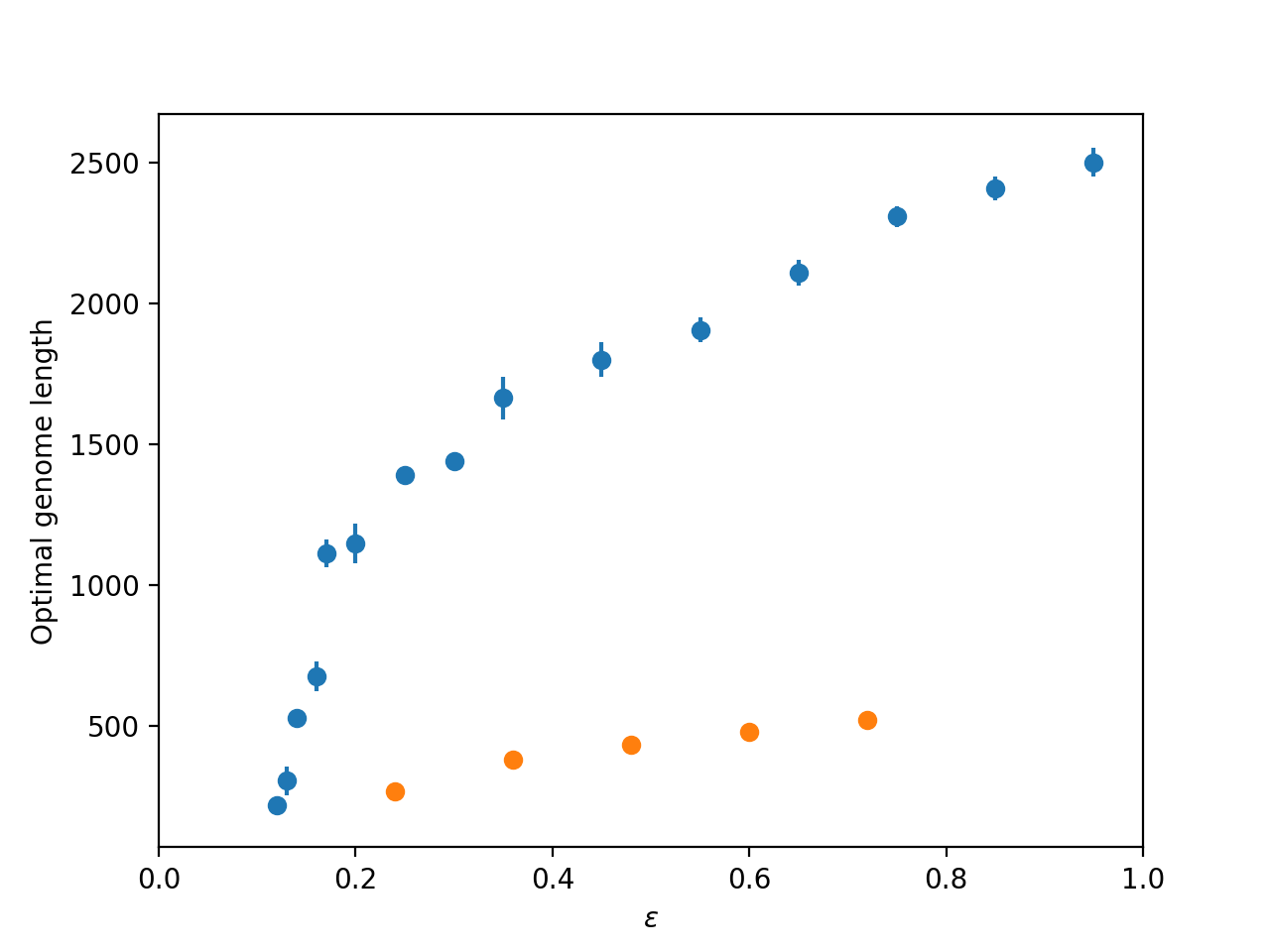}
		\caption{Optimal length of the genome as a function of the strength of the interaction between N-terminal and genome $\varepsilon$ for a $T=3$ capsid (blue curve) and inner radius $R=21$ nm and a $T=1$ capsid (orange curve) and inner radius $R=12$ nm the $n=5$ also for both capsids. We expected the linear relation between them.}	
		\centering 
		\label{charge_density}
	\end{center}
\end{figure}

In our model, the strength of the Lennard-Jones interaction between the beads of the polymer chain and those of the N-terminal chains reflects the influence of the charge density of these chains or the ionic strength of the buffer solution. According to theory, we expect a linear relationship between the interaction strength and the optimal chain length for sufficiently weak interactions. Our findings, shown in Fig.~\ref{charge_density}, confirm this expectation but also reveal that for stronger interaction strengths, the dependence of the optimal genome length on $\varepsilon$ becomes weaker than linear. The figure also suggests that a minimum interaction strength is required for the genome to be encapsulated. This is perhaps not entirely surprising, as it is well known that a minimum attraction strength is required for a polymer to adsorb onto a surface. This minimum attraction strength depends on the polymer length, albeit only weakly~\cite{Piculell1995}. The simple Flory theory, with which we make our comparisons, does not account for this. The required minimum interaction strength is related to the loss of translational entropy of the chain~\cite{rubinstein2003polymer}.

\subsection{Effect of the radius of a capsid}

\begin{figure}[h]
\begin{center}\includegraphics[width=8cm]{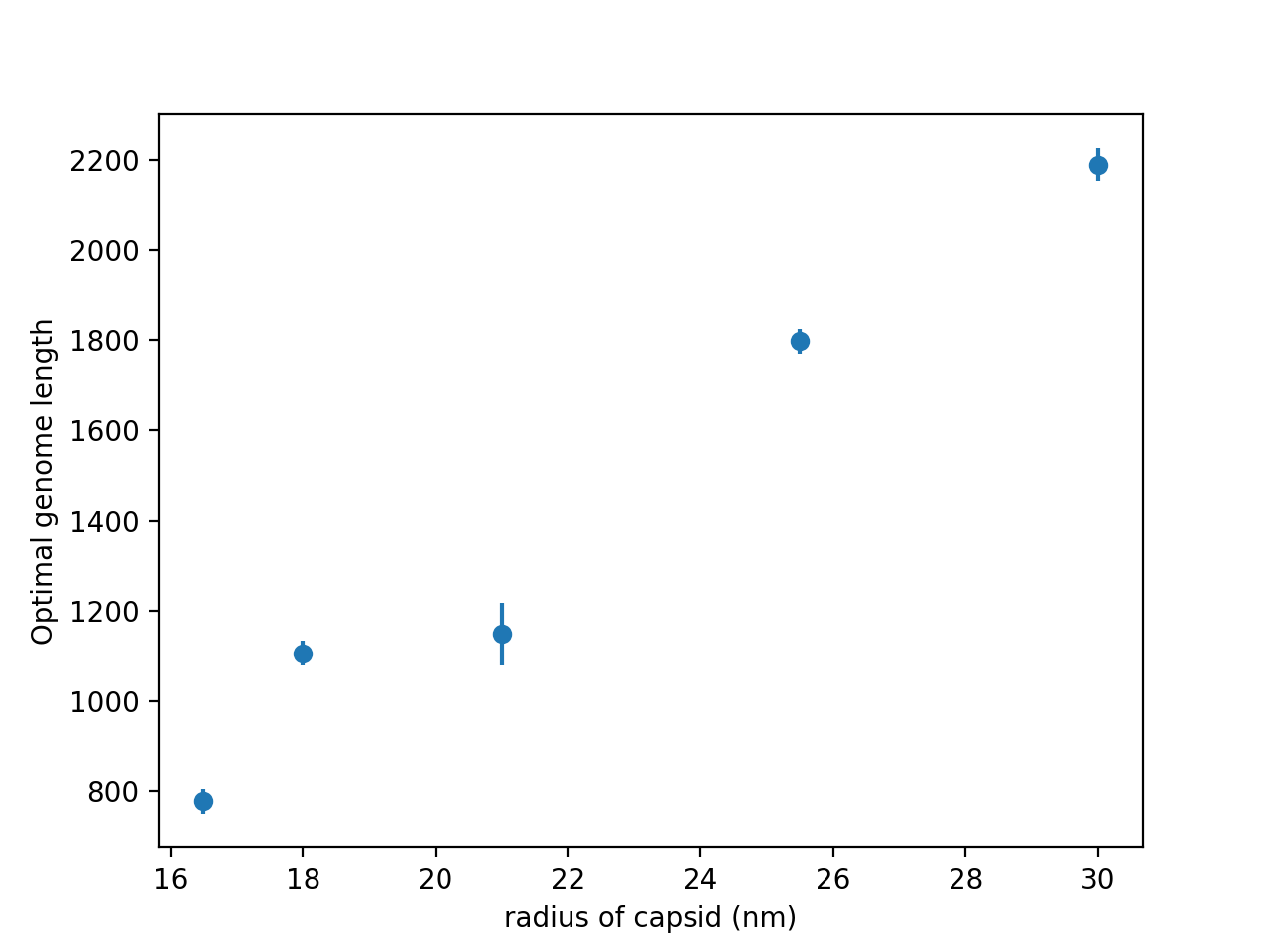}\caption{Optimal genome length as a function of the radius of a $T=3$ capsid for a given interaction energy $\varepsilon = 0.2$. The length of the N-terminals is $n=5$ beads.
}	
		\centering 
		\label{radius}
	\end{center}
\end{figure}

As already mentioned, the radius $R$ of a capsid is directly related to its $T$ number, with $R \propto T^{1/2}$. However, there is considerable variation even within a single T number~\cite{bovzivc2017varieties}. According to Flory’s theory~\cite{van2013impact}, the optimal length of the encapsulated genome should not scale with the radius if the number of N-terminals is kept constant and the focus remains on a single T number.
The results of our simulations, shown in Fig.\ref{radius}, indicate that as the radius increases, the encapsulated genome length increases proportionally, presumably due to the additional space available, which reduces the loss of conformational entropy of the genome. Perlmutter \textit{et al.} also performed computer simulations on genome packaging and found that packing increases with the radius of the capsid as $M \propto R^{1.6}$\cite{perlmutter2013viral}, rather than the $M \propto R^0$ scaling predicted by Flory's theory~\cite{van2013impact} and the polymer density functional theory~\cite{van2005electrostatics}. We observe a somewhat smaller scaling exponent in our simulations. Clearly, Flory’s theory does not capture this behavior~\cite{van2013impact}, nor does the density functional theory, where the interaction with the N-terminals is averaged over the entire surface of the shell~\cite{van2005electrostatics}. For very large radii, where the problem resembles that of a flat surface with localized regions of attraction, we expect the optimal genome length to level off, though this effect appears to occur for capsid sizes much larger than those we investigate.

\subsection{Impact of a charge distribution} 
\begin{figure}[h]
	\begin{center}
		\includegraphics[width=6cm]{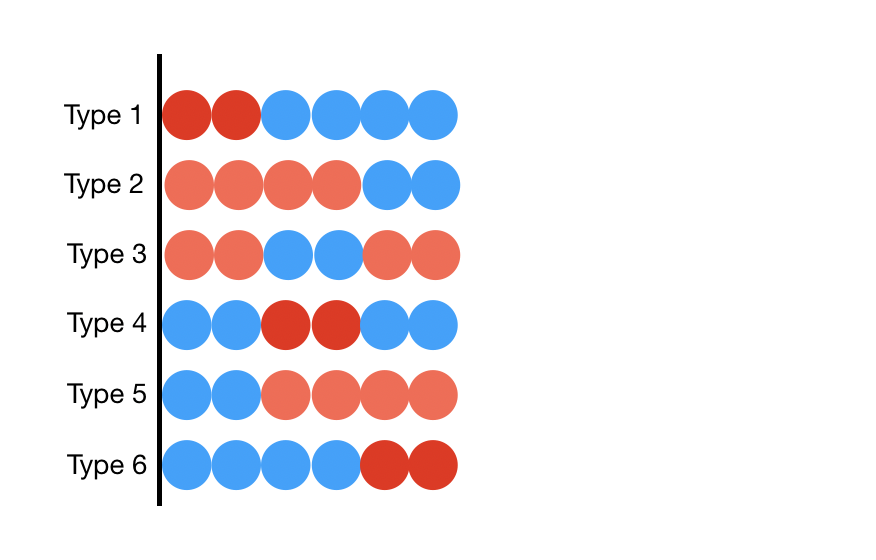}
		\caption{Model of six distinct charge configurations along the N-terminals. A positive charge is represented by the red or orange circle, while a neutral charge is displayed by the blue circle.The red and orange circles represent different charge densities, indicated by varying values of $\varepsilon$ in our model. The total charge is assumed to be fixed, which corresponds to a constant total interaction energy of $n \times \varepsilon = 1.24$, where $n$ denotes the number of charged beads along the chain.}	
		\centering 
		\label{fig7}
	\end{center}
\end{figure}

Typically, the positively charged moieties on the N-terminals are not evenly distributed along the primary sequence but are clustered in specific regions~\cite{bovzivc2017varieties}. Recent experiments point out that the kind of basic amino acid as well as the position along the backbone of the N-terminals both have a large effect on the encapsulated mass of genome~\cite{ni2012examination}. In our simulations, we investigated two types of N-terminals: one with two attractive beads (representing charged amino acids) and four neutral beads with only excluded volume interactions, and the other with four attractive beads and two neutral beads, also with excluded volume interactions only. By varying the positions of the charged beads, we created six different structures. For the second type, the attraction between an attractive bead and a chain bead is half the strength of the interaction in the first type, ensuring that the total attractive interaction remains constant. See Figure~\ref{fig7} for a pictorial representation of our six N-terminal model types.

\begin{figure}[h]
	\begin{center}
\includegraphics[width=8cm]{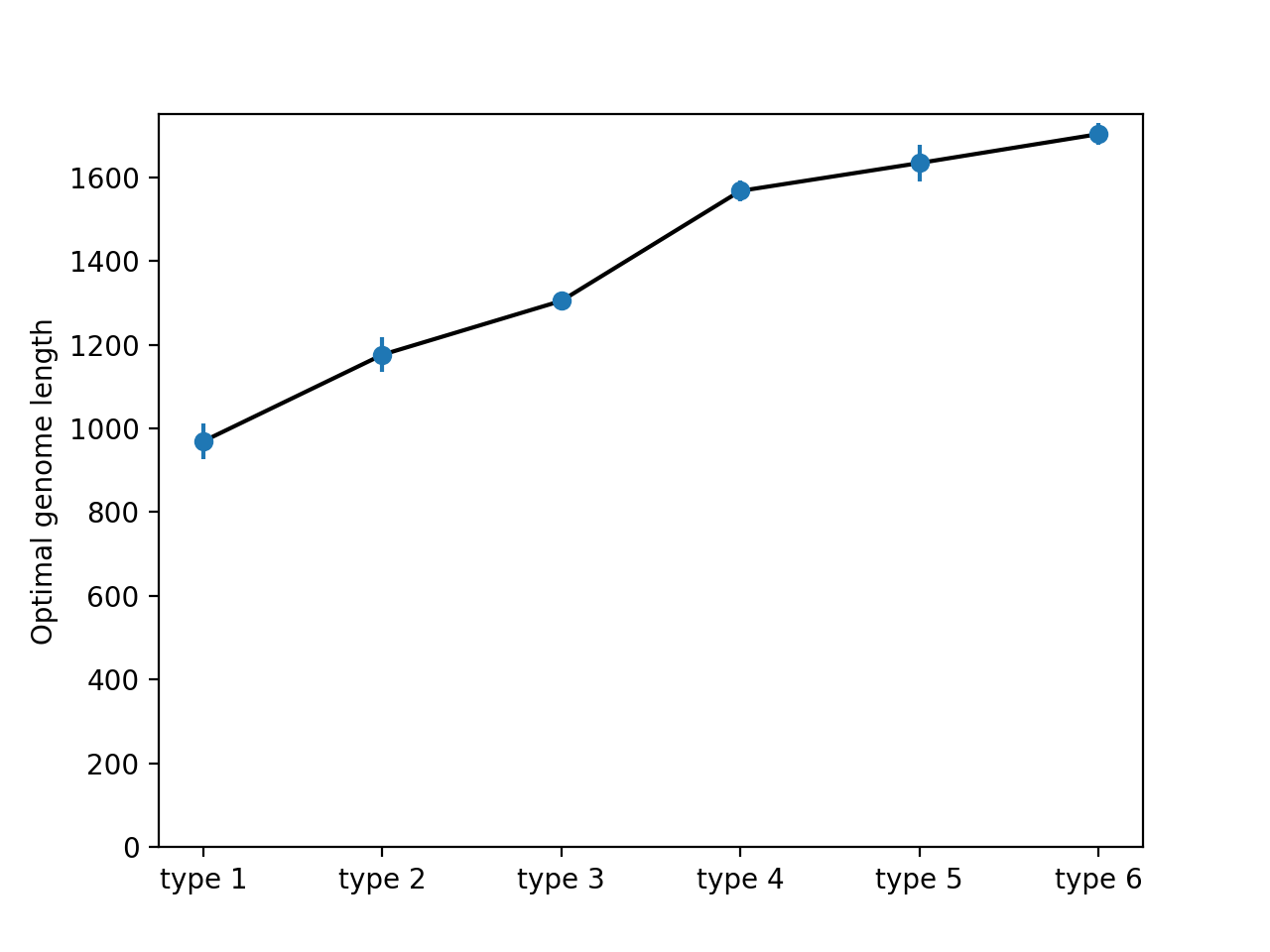}
		\caption{Optimal length of the genome for a $T=3$ capsid of radius $R=21$ $nm$ for the different N-terminal models shown in Fig.~\ref{fig7}. The total charge is fixed, which translates to a total interaction energy $n \times \varepsilon = 1.24$ that is constant where $n$ here represents the number of charged beads along the chain.}	
		\centering 
		\label{fig5}
	\end{center}
\end{figure}

Figure~\ref{fig5} shows that the optimal encapsulated genome length increases from type 1 to type 6. Comparing types 1 and 6 shows that moving the two attractive beads away from the capsid wall increases the encapsulated genome by more than 60 percent. This is perhaps not all that surprising, given that the presence of a repulsive capsid wall must have a stronger effect on the polymer configurations if the attractive beads are closer by~\cite{Huang2021}. Doubling the number of charged beads while halving the strength of attractive interaction also causes the adsorbed polymer to move away from the surface. Specifically, going from type 1 to type 2 results in the third and fourth charged beads being positioned further from the capsid wall. Comparing type 2 with types 3 and 5, where two and four beads are located nearer the end of the N-terminal, shows an increase in the number of encapsulated chain segments. Interestingly, replacing two full-strength beads at the end of the N-terminal with four half-strength beads has relatively little effect. This can be expected due to the increased distance from the capsid wall. 

\begin{figure}[h]
	\begin{center}
\includegraphics[width=8cm]{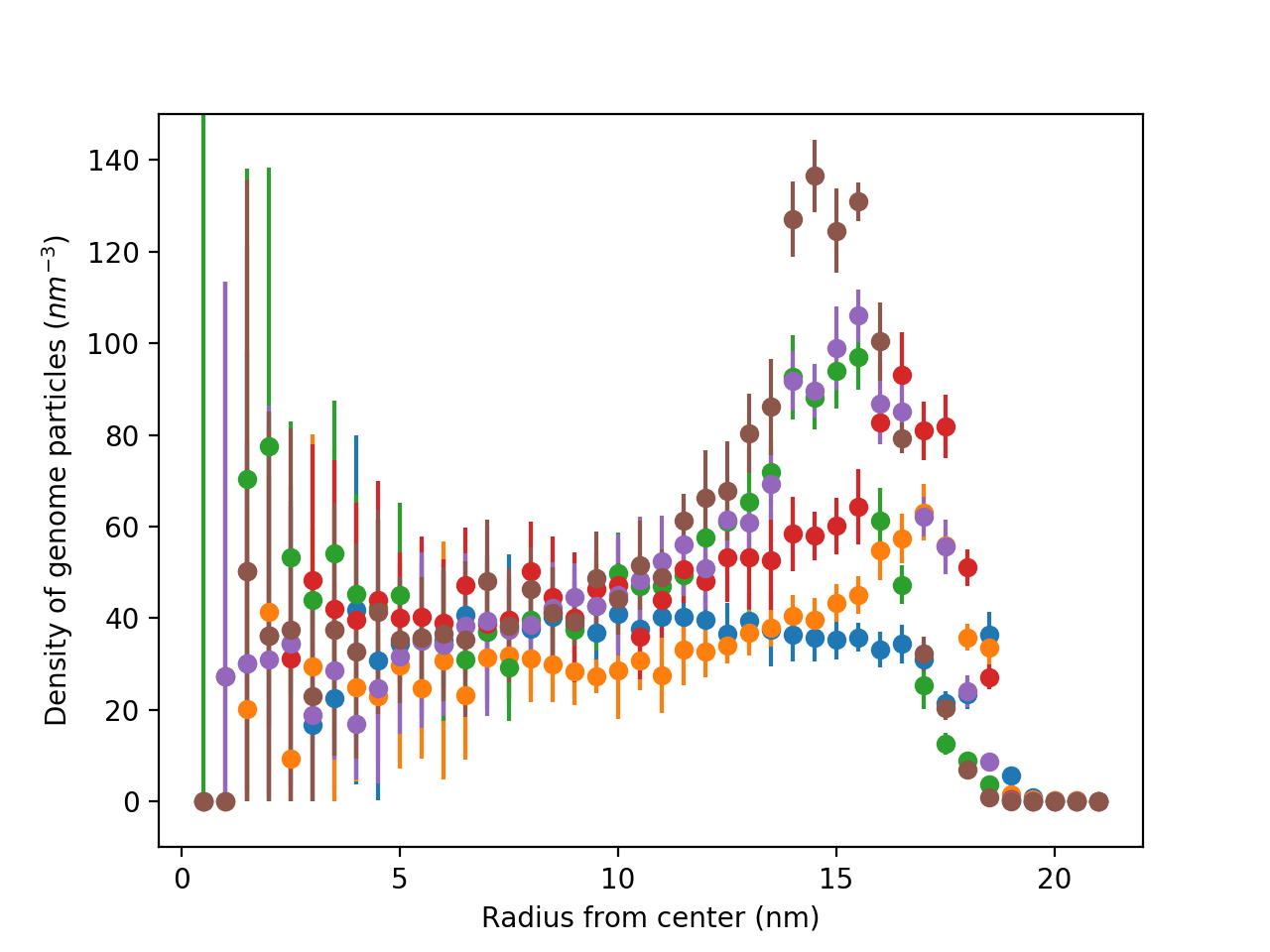}
		\caption{Density of the genomic beads  (in units of nm$^{-3}$) as a function of the radial distance from the center of the capsid for a $T=3$ capsid with an inner radius of $R=21$ nm, corresponding to the different N-terminal models shown in Fig.~\ref{fig7}. The total charge is fixed, resulting in a constant total interaction energy of $n \varepsilon = 1.24$, where $n$ now denotes the number of charged beads along the chain. Types 1 through 6 are represented by blue, orange, green, red, purple, and brown symbols, respectively.}	
		\centering 
		\label{diff_density_pos}
	\end{center}
\end{figure}

This is confirmed by the genome density profiles that we plotted in Fig.~\ref{diff_density_pos}, where we see that the density decreases at a radial distance from the center of the cavity well before 19 $nm$, which is roughly were the interaction between bead and wall particles becomes strongly repulsive. The peak positions match the positions of the charged groups, where the peak height correlates positively with the optimal encapsulated length of the cargo. N-terminal type 6 has the largest peak and appears to be the most localized, with a relatively narrow distribution around the charged groups. We also notice that the decay length of the density towards the center of the cavity is comparable across all types. In the region of the cavity where the N-terminals cannot reach, the density of chain segments is flat and shows little variation between different N-terminal types.  A narrow region in the very center of the cavity is virtually devoid of segments with mean densities that have the largest spread.  Interestingly, the spread in the low-density region near the wall of the capsid is quite small. This is probably the effect of the harshly repulsive interaction between cargo and wall beads.

\section{Conclusion}
We studied the optimal length of the encapsulated genome in relation to the length and structure of the N-terminals in $T=1$ and $T=3$ capsids using computer simulations. Rather than explicitly modeling the Coulomb interactions between the genome and the RNA binding domains, we employed attractive Lennard-Jones potentials as a proxy for these interactions. The primary motivation for this approach is that introducing explicit charges on the genome and RNA binding domains, along with mobile ions in the simulation volume, is computationally very expensive. Moreover, due to Donnan effects, the ion concentrations inside and outside the capsid may differ~\cite{muhren2023electrostatic}, requiring substantial equilibration times. Accounting for specific ion effects in a meaningful way also introduces additional complexity~\cite{Ninham2012}. By using model potentials, we effectively circumvent these challenges while maintaining computational feasibility.

We investigated the impact of several factors on packaging efficiency, including the strength of the interaction between the N-terminal RNA binding domains and a polymer chain representing the genome, the length of the N-terminals, the capsid radius, and the localization of interactions between the N-terminals and the polymer chain. We find that the optimal genome length increases with the interaction strength but does so sub-linearly across the entire range we tested. Similarly, the optimal genome length increases with the length of the N-terminals, even when the total attraction strength (representing a constant number of charges) remains fixed. This observation is consistent with the field-theoretical results of Dong et al.\cite{dong2020effect}. Overall, while our findings are in qualitative agreement with the predictions of the simple Flory theory\cite{van2013impact}, they exhibit notable quantitative differences.

In agreement with the field-theoretical work of Dong et al.~\cite{dong2020effect}, we find that the optimal length of the encapsulated genome depends on the spatial distribution of the charges (attraction sites) along the N-terminal. When the charges are positioned near the capsid wall, genome packaging is reduced. We attribute this to the influence of a rigid wall, which constrains polymer conformations in its vicinity. Additionally, increasing the capsid radius while keeping the number of charges constant (i.e., for a fixed T number) leads to an increase in the optimal genome length. This effect can be attributed to the additional space available, which expands the conformational freedom of the genome.

When comparing T=3 and T=1 capsids, two key parameters must be considered: the total charge on the capsid and the capsid radius. First, the increase in the number of proteins from T=1 to T=3 raises the total count from 60 to 180, which means the total number of charges on the N-terminals also triples. Given the known linear relationship between total charge and optimal genome length, we would expect the encapsulated genome length to increase threefold. However, it is worth noting that virtually all data supporting this relationship pertain to T=3 particles~\cite{belyi2006electrostatic,bovzivc2017varieties,hu2008electrostatic}.

In our simulations, we observe an increase closer to a factor of four to five, indicating that the capsid radius also plays a significant role. We propose that the localization of charges on the N-terminals introduces a radius dependence that is overlooked in highly coarse-grained models.

\acknowledgments{The authors acknowledge support from NSF DMR-2131963 and the University of California Multicampus Research Programs and Initiatives (Grant No. M21PR3267).}

\bibliography{main} 

\begin{thebibliography}{86}%
\makeatletter
\providecommand \@ifxundefined [1]{%
 \@ifx{#1\undefined}
}%
\providecommand \@ifnum [1]{%
 \ifnum #1\expandafter \@firstoftwo
 \else \expandafter \@secondoftwo
 \fi
}%
\providecommand \@ifx [1]{%
 \ifx #1\expandafter \@firstoftwo
 \else \expandafter \@secondoftwo
 \fi
}%
\providecommand \natexlab [1]{#1}%
\providecommand \enquote  [1]{``#1''}%
\providecommand \bibnamefont  [1]{#1}%
\providecommand \bibfnamefont [1]{#1}%
\providecommand \citenamefont [1]{#1}%
\providecommand \href@noop [0]{\@secondoftwo}%
\providecommand \href [0]{\begingroup \@sanitize@url \@href}%
\providecommand \@href[1]{\@@startlink{#1}\@@href}%
\providecommand \@@href[1]{\endgroup#1\@@endlink}%
\providecommand \@sanitize@url [0]{\catcode `\\12\catcode `\$12\catcode `\&12\catcode `\#12\catcode `\^12\catcode `\_12\catcode `\%12\relax}%
\providecommand \@@startlink[1]{}%
\providecommand \@@endlink[0]{}%
\providecommand \url  [0]{\begingroup\@sanitize@url \@url }%
\providecommand \@url [1]{\endgroup\@href {#1}{\urlprefix }}%
\providecommand \urlprefix  [0]{URL }%
\providecommand \Eprint [0]{\href }%
\providecommand \doibase [0]{https://doi.org/}%
\providecommand \selectlanguage [0]{\@gobble}%
\providecommand \bibinfo  [0]{\@secondoftwo}%
\providecommand \bibfield  [0]{\@secondoftwo}%
\providecommand \translation [1]{[#1]}%
\providecommand \BibitemOpen [0]{}%
\providecommand \bibitemStop [0]{}%
\providecommand \bibitemNoStop [0]{.\EOS\space}%
\providecommand \EOS [0]{\spacefactor3000\relax}%
\providecommand \BibitemShut  [1]{\csname bibitem#1\endcsname}%
\let\auto@bib@innerbib\@empty
\bibitem [{\citenamefont {Comas-Garcia}\ \emph {et~al.}(2012)\citenamefont {Comas-Garcia}, \citenamefont {Cadena-Nava}, \citenamefont {Rao}, \citenamefont {Knobler},\ and\ \citenamefont {Gelbart}}]{comas2012vitro}%
  \BibitemOpen
  \bibfield  {author} {\bibinfo {author} {\bibfnamefont {M.}~\bibnamefont {Comas-Garcia}}, \bibinfo {author} {\bibfnamefont {R.~D.}\ \bibnamefont {Cadena-Nava}}, \bibinfo {author} {\bibfnamefont {A.}~\bibnamefont {Rao}}, \bibinfo {author} {\bibfnamefont {C.~M.}\ \bibnamefont {Knobler}},\ and\ \bibinfo {author} {\bibfnamefont {W.~M.}\ \bibnamefont {Gelbart}},\ }\bibfield  {title} {\bibinfo {title} {In vitro quantification of the relative packaging efficiencies of single-stranded {RNA} molecules by viral capsid protein},\ }\href@noop {} {\bibfield  {journal} {\bibinfo  {journal} {Journal of virology}\ }\textbf {\bibinfo {volume} {86}},\ \bibinfo {pages} {12271} (\bibinfo {year} {2012})}\BibitemShut {NoStop}%
\bibitem [{\citenamefont {Vaupoti{\v{c}}}\ \emph {et~al.}(2023)\citenamefont {Vaupoti{\v{c}}}, \citenamefont {Rosa}, \citenamefont {Podgornik}, \citenamefont {Tubiana},\ and\ \citenamefont {Bo{\v{z}}i{\v{c}}}}]{vaupotivc2023viral}%
  \BibitemOpen
  \bibfield  {author} {\bibinfo {author} {\bibfnamefont {D.}~\bibnamefont {Vaupoti{\v{c}}}}, \bibinfo {author} {\bibfnamefont {A.}~\bibnamefont {Rosa}}, \bibinfo {author} {\bibfnamefont {R.}~\bibnamefont {Podgornik}}, \bibinfo {author} {\bibfnamefont {L.}~\bibnamefont {Tubiana}},\ and\ \bibinfo {author} {\bibfnamefont {A.}~\bibnamefont {Bo{\v{z}}i{\v{c}}}},\ }\bibfield  {title} {\bibinfo {title} {Viral {RNA} as a branched polymer},\ }in\ \href@noop {} {\emph {\bibinfo {booktitle} {Physical Virology: From the State-of-the-Art Research to the Future of Applied Virology}}}\ (\bibinfo  {publisher} {Springer},\ \bibinfo {year} {2023})\ pp.\ \bibinfo {pages} {1--26}\BibitemShut {NoStop}%
\bibitem [{\citenamefont {Hagan}\ and\ \citenamefont {Zandi}(2016)}]{hagan2016recent}%
  \BibitemOpen
  \bibfield  {author} {\bibinfo {author} {\bibfnamefont {M.~F.}\ \bibnamefont {Hagan}}\ and\ \bibinfo {author} {\bibfnamefont {R.}~\bibnamefont {Zandi}},\ }\bibfield  {title} {\bibinfo {title} {Recent advances in coarse-grained modeling of virus assembly},\ }\href@noop {} {\bibfield  {journal} {\bibinfo  {journal} {Current opinion in virology}\ }\textbf {\bibinfo {volume} {18}},\ \bibinfo {pages} {36} (\bibinfo {year} {2016})}\BibitemShut {NoStop}%
\bibitem [{\citenamefont {Ning}\ \emph {et~al.}(2016)\citenamefont {Ning}, \citenamefont {Erdemci-Tandogan}, \citenamefont {Yufenyuy}, \citenamefont {Wagner}, \citenamefont {Himes}, \citenamefont {Zhao}, \citenamefont {Aiken}, \citenamefont {Zandi},\ and\ \citenamefont {Zhang}}]{ning2016vitro}%
  \BibitemOpen
  \bibfield  {author} {\bibinfo {author} {\bibfnamefont {J.}~\bibnamefont {Ning}}, \bibinfo {author} {\bibfnamefont {G.}~\bibnamefont {Erdemci-Tandogan}}, \bibinfo {author} {\bibfnamefont {E.~L.}\ \bibnamefont {Yufenyuy}}, \bibinfo {author} {\bibfnamefont {J.}~\bibnamefont {Wagner}}, \bibinfo {author} {\bibfnamefont {B.~A.}\ \bibnamefont {Himes}}, \bibinfo {author} {\bibfnamefont {G.}~\bibnamefont {Zhao}}, \bibinfo {author} {\bibfnamefont {C.}~\bibnamefont {Aiken}}, \bibinfo {author} {\bibfnamefont {R.}~\bibnamefont {Zandi}},\ and\ \bibinfo {author} {\bibfnamefont {P.}~\bibnamefont {Zhang}},\ }\bibfield  {title} {\bibinfo {title} {In vitro protease cleavage and computer simulations reveal the {HIV}-1 capsid maturation pathway},\ }\href@noop {} {\bibfield  {journal} {\bibinfo  {journal} {Nature communications}\ }\textbf {\bibinfo {volume} {7}},\ \bibinfo {pages} {13689} (\bibinfo {year} {2016})}\BibitemShut {NoStop}%
\bibitem [{\citenamefont {Beren}\ \emph {et~al.}(2017)\citenamefont {Beren}, \citenamefont {Dreesens}, \citenamefont {Liu}, \citenamefont {Knobler},\ and\ \citenamefont {Gelbart}}]{beren2017effect}%
  \BibitemOpen
  \bibfield  {author} {\bibinfo {author} {\bibfnamefont {C.}~\bibnamefont {Beren}}, \bibinfo {author} {\bibfnamefont {L.~L.}\ \bibnamefont {Dreesens}}, \bibinfo {author} {\bibfnamefont {K.~N.}\ \bibnamefont {Liu}}, \bibinfo {author} {\bibfnamefont {C.~M.}\ \bibnamefont {Knobler}},\ and\ \bibinfo {author} {\bibfnamefont {W.~M.}\ \bibnamefont {Gelbart}},\ }\bibfield  {title} {\bibinfo {title} {The effect of {RNA} secondary structure on the self-assembly of viral capsids},\ }\href@noop {} {\bibfield  {journal} {\bibinfo  {journal} {Biophysical journal}\ }\textbf {\bibinfo {volume} {113}},\ \bibinfo {pages} {339} (\bibinfo {year} {2017})}\BibitemShut {NoStop}%
\bibitem [{\citenamefont {Perlmutter}\ and\ \citenamefont {Hagan}(2015{\natexlab{a}})}]{perlmutter2015role}%
  \BibitemOpen
  \bibfield  {author} {\bibinfo {author} {\bibfnamefont {J.~D.}\ \bibnamefont {Perlmutter}}\ and\ \bibinfo {author} {\bibfnamefont {M.~F.}\ \bibnamefont {Hagan}},\ }\bibfield  {title} {\bibinfo {title} {The role of packaging sites in efficient and specific virus assembly},\ }\href@noop {} {\bibfield  {journal} {\bibinfo  {journal} {Journal of molecular biology}\ }\textbf {\bibinfo {volume} {427}},\ \bibinfo {pages} {2451} (\bibinfo {year} {2015}{\natexlab{a}})}\BibitemShut {NoStop}%
\bibitem [{\citenamefont {Stockley}\ \emph {et~al.}(2013)\citenamefont {Stockley}, \citenamefont {Twarock}, \citenamefont {Bakker}, \citenamefont {Barker}, \citenamefont {Borodavka}, \citenamefont {Dykeman}, \citenamefont {Ford}, \citenamefont {Pearson}, \citenamefont {Phillips}, \citenamefont {Ranson} \emph {et~al.}}]{stockley2013packaging}%
  \BibitemOpen
  \bibfield  {author} {\bibinfo {author} {\bibfnamefont {P.~G.}\ \bibnamefont {Stockley}}, \bibinfo {author} {\bibfnamefont {R.}~\bibnamefont {Twarock}}, \bibinfo {author} {\bibfnamefont {S.~E.}\ \bibnamefont {Bakker}}, \bibinfo {author} {\bibfnamefont {A.~M.}\ \bibnamefont {Barker}}, \bibinfo {author} {\bibfnamefont {A.}~\bibnamefont {Borodavka}}, \bibinfo {author} {\bibfnamefont {E.}~\bibnamefont {Dykeman}}, \bibinfo {author} {\bibfnamefont {R.~J.}\ \bibnamefont {Ford}}, \bibinfo {author} {\bibfnamefont {A.~R.}\ \bibnamefont {Pearson}}, \bibinfo {author} {\bibfnamefont {S.~E.}\ \bibnamefont {Phillips}}, \bibinfo {author} {\bibfnamefont {N.~A.}\ \bibnamefont {Ranson}}, \emph {et~al.},\ }\bibfield  {title} {\bibinfo {title} {Packaging signals in single-stranded {RNA} viruses: nature’s alternative to a purely electrostatic assembly mechanism},\ }\href@noop {} {\bibfield  {journal} {\bibinfo  {journal} {Journal of biological physics}\ }\textbf {\bibinfo {volume} {39}},\ \bibinfo {pages} {277} (\bibinfo {year}
  {2013})}\BibitemShut {NoStop}%
\bibitem [{\citenamefont {Panahandeh}\ \emph {et~al.}(2022)\citenamefont {Panahandeh}, \citenamefont {Li}, \citenamefont {Dragnea},\ and\ \citenamefont {Zandi}}]{panahandeh2022virus}%
  \BibitemOpen
  \bibfield  {author} {\bibinfo {author} {\bibfnamefont {S.}~\bibnamefont {Panahandeh}}, \bibinfo {author} {\bibfnamefont {S.}~\bibnamefont {Li}}, \bibinfo {author} {\bibfnamefont {B.}~\bibnamefont {Dragnea}},\ and\ \bibinfo {author} {\bibfnamefont {R.}~\bibnamefont {Zandi}},\ }\bibfield  {title} {\bibinfo {title} {Virus assembly pathways inside a host cell},\ }\href@noop {} {\bibfield  {journal} {\bibinfo  {journal} {ACS nano}\ }\textbf {\bibinfo {volume} {16}},\ \bibinfo {pages} {317} (\bibinfo {year} {2022})}\BibitemShut {NoStop}%
\bibitem [{\citenamefont {Cadena-Nava}\ \emph {et~al.}(2011)\citenamefont {Cadena-Nava}, \citenamefont {Hu}, \citenamefont {Garmann}, \citenamefont {Ng}, \citenamefont {Zelikin}, \citenamefont {Knobler},\ and\ \citenamefont {Gelbart}}]{cadena2011exploiting}%
  \BibitemOpen
  \bibfield  {author} {\bibinfo {author} {\bibfnamefont {R.~D.}\ \bibnamefont {Cadena-Nava}}, \bibinfo {author} {\bibfnamefont {Y.}~\bibnamefont {Hu}}, \bibinfo {author} {\bibfnamefont {R.~F.}\ \bibnamefont {Garmann}}, \bibinfo {author} {\bibfnamefont {B.}~\bibnamefont {Ng}}, \bibinfo {author} {\bibfnamefont {A.~N.}\ \bibnamefont {Zelikin}}, \bibinfo {author} {\bibfnamefont {C.~M.}\ \bibnamefont {Knobler}},\ and\ \bibinfo {author} {\bibfnamefont {W.~M.}\ \bibnamefont {Gelbart}},\ }\bibfield  {title} {\bibinfo {title} {Exploiting fluorescent polymers to probe the self-assembly of virus-like particles},\ }\href@noop {} {\bibfield  {journal} {\bibinfo  {journal} {The Journal of Physical Chemistry B}\ }\textbf {\bibinfo {volume} {115}},\ \bibinfo {pages} {2386} (\bibinfo {year} {2011})}\BibitemShut {NoStop}%
\bibitem [{\citenamefont {Sun}\ \emph {et~al.}(2007)\citenamefont {Sun}, \citenamefont {DuFort}, \citenamefont {Daniel}, \citenamefont {Murali}, \citenamefont {Chen}, \citenamefont {Gopinath}, \citenamefont {Stein}, \citenamefont {De}, \citenamefont {Rotello}, \citenamefont {Holzenburg} \emph {et~al.}}]{sun2007core}%
  \BibitemOpen
  \bibfield  {author} {\bibinfo {author} {\bibfnamefont {J.}~\bibnamefont {Sun}}, \bibinfo {author} {\bibfnamefont {C.}~\bibnamefont {DuFort}}, \bibinfo {author} {\bibfnamefont {M.-C.}\ \bibnamefont {Daniel}}, \bibinfo {author} {\bibfnamefont {A.}~\bibnamefont {Murali}}, \bibinfo {author} {\bibfnamefont {C.}~\bibnamefont {Chen}}, \bibinfo {author} {\bibfnamefont {K.}~\bibnamefont {Gopinath}}, \bibinfo {author} {\bibfnamefont {B.}~\bibnamefont {Stein}}, \bibinfo {author} {\bibfnamefont {M.}~\bibnamefont {De}}, \bibinfo {author} {\bibfnamefont {V.~M.}\ \bibnamefont {Rotello}}, \bibinfo {author} {\bibfnamefont {A.}~\bibnamefont {Holzenburg}}, \emph {et~al.},\ }\bibfield  {title} {\bibinfo {title} {Core-controlled polymorphism in virus-like particles},\ }\href@noop {} {\bibfield  {journal} {\bibinfo  {journal} {Proceedings of the National Academy of Sciences}\ }\textbf {\bibinfo {volume} {104}},\ \bibinfo {pages} {1354} (\bibinfo {year} {2007})}\BibitemShut {NoStop}%
\bibitem [{\citenamefont {Panahandeh}\ \emph {et~al.}(2020)\citenamefont {Panahandeh}, \citenamefont {Li}, \citenamefont {Marichal}, \citenamefont {Leite~Rubim}, \citenamefont {Tresset},\ and\ \citenamefont {Zandi}}]{panahandeh2020virus}%
  \BibitemOpen
  \bibfield  {author} {\bibinfo {author} {\bibfnamefont {S.}~\bibnamefont {Panahandeh}}, \bibinfo {author} {\bibfnamefont {S.}~\bibnamefont {Li}}, \bibinfo {author} {\bibfnamefont {L.}~\bibnamefont {Marichal}}, \bibinfo {author} {\bibfnamefont {R.}~\bibnamefont {Leite~Rubim}}, \bibinfo {author} {\bibfnamefont {G.}~\bibnamefont {Tresset}},\ and\ \bibinfo {author} {\bibfnamefont {R.}~\bibnamefont {Zandi}},\ }\bibfield  {title} {\bibinfo {title} {How a virus circumvents energy barriers to form symmetric shells},\ }\href@noop {} {\bibfield  {journal} {\bibinfo  {journal} {ACS nano}\ }\textbf {\bibinfo {volume} {14}},\ \bibinfo {pages} {3170} (\bibinfo {year} {2020})}\BibitemShut {NoStop}%
\bibitem [{\citenamefont {Zandi}\ \emph {et~al.}(2020)\citenamefont {Zandi}, \citenamefont {Dragnea}, \citenamefont {Travesset},\ and\ \citenamefont {Podgornik}}]{zandi2020virus}%
  \BibitemOpen
  \bibfield  {author} {\bibinfo {author} {\bibfnamefont {R.}~\bibnamefont {Zandi}}, \bibinfo {author} {\bibfnamefont {B.}~\bibnamefont {Dragnea}}, \bibinfo {author} {\bibfnamefont {A.}~\bibnamefont {Travesset}},\ and\ \bibinfo {author} {\bibfnamefont {R.}~\bibnamefont {Podgornik}},\ }\bibfield  {title} {\bibinfo {title} {On virus growth and form},\ }\href@noop {} {\bibfield  {journal} {\bibinfo  {journal} {Physics Reports}\ }\textbf {\bibinfo {volume} {847}},\ \bibinfo {pages} {1} (\bibinfo {year} {2020})}\BibitemShut {NoStop}%
\bibitem [{\citenamefont {Johnson}\ and\ \citenamefont {Chiu}(2000)}]{johnson2000structures}%
  \BibitemOpen
  \bibfield  {author} {\bibinfo {author} {\bibfnamefont {J.~E.}\ \bibnamefont {Johnson}}\ and\ \bibinfo {author} {\bibfnamefont {W.}~\bibnamefont {Chiu}},\ }\bibfield  {title} {\bibinfo {title} {Structures of virus and virus-like particles},\ }\href@noop {} {\bibfield  {journal} {\bibinfo  {journal} {Current opinion in structural biology}\ }\textbf {\bibinfo {volume} {10}},\ \bibinfo {pages} {229} (\bibinfo {year} {2000})}\BibitemShut {NoStop}%
\bibitem [{\citenamefont {Cad{\`e}ne}\ \emph {et~al.}(2014)\citenamefont {Cad{\`e}ne}, \citenamefont {Berceau}, \citenamefont {Fouch{\'e}}, \citenamefont {Battesti},\ and\ \citenamefont {Rizzo}}]{cadene2014vacuum}%
  \BibitemOpen
  \bibfield  {author} {\bibinfo {author} {\bibfnamefont {A.}~\bibnamefont {Cad{\`e}ne}}, \bibinfo {author} {\bibfnamefont {P.}~\bibnamefont {Berceau}}, \bibinfo {author} {\bibfnamefont {M.}~\bibnamefont {Fouch{\'e}}}, \bibinfo {author} {\bibfnamefont {R.}~\bibnamefont {Battesti}},\ and\ \bibinfo {author} {\bibfnamefont {C.}~\bibnamefont {Rizzo}},\ }\bibfield  {title} {\bibinfo {title} {Vacuum magnetic linear birefringence using pulsed fields: status of the {BMV} experiment},\ }\href@noop {} {\bibfield  {journal} {\bibinfo  {journal} {The European Physical Journal D}\ }\textbf {\bibinfo {volume} {68}},\ \bibinfo {pages} {1} (\bibinfo {year} {2014})}\BibitemShut {NoStop}%
\bibitem [{\citenamefont {Battesti}\ \emph {et~al.}(2008)\citenamefont {Battesti}, \citenamefont {Pinto Da~Souza}, \citenamefont {Batut}, \citenamefont {Robilliard}, \citenamefont {Bailly}, \citenamefont {Michel}, \citenamefont {Nardone}, \citenamefont {Pinard}, \citenamefont {Portugall}, \citenamefont {Tr{\'e}nec} \emph {et~al.}}]{battesti2008bmv}%
  \BibitemOpen
  \bibfield  {author} {\bibinfo {author} {\bibfnamefont {R.}~\bibnamefont {Battesti}}, \bibinfo {author} {\bibfnamefont {B.}~\bibnamefont {Pinto Da~Souza}}, \bibinfo {author} {\bibfnamefont {S.}~\bibnamefont {Batut}}, \bibinfo {author} {\bibfnamefont {C.}~\bibnamefont {Robilliard}}, \bibinfo {author} {\bibfnamefont {G.}~\bibnamefont {Bailly}}, \bibinfo {author} {\bibfnamefont {C.}~\bibnamefont {Michel}}, \bibinfo {author} {\bibfnamefont {M.}~\bibnamefont {Nardone}}, \bibinfo {author} {\bibfnamefont {L.}~\bibnamefont {Pinard}}, \bibinfo {author} {\bibfnamefont {O.}~\bibnamefont {Portugall}}, \bibinfo {author} {\bibfnamefont {G.}~\bibnamefont {Tr{\'e}nec}}, \emph {et~al.},\ }\bibfield  {title} {\bibinfo {title} {The {BMV} experiment: a novel apparatus to study the propagation of light in a transverse magnetic field},\ }\href@noop {} {\bibfield  {journal} {\bibinfo  {journal} {The European Physical Journal D}\ }\textbf {\bibinfo {volume} {46}},\ \bibinfo {pages} {323} (\bibinfo {year} {2008})}\BibitemShut
  {NoStop}%
\bibitem [{\citenamefont {Van~der Graaf}\ \emph {et~al.}(1991)\citenamefont {Van~der Graaf}, \citenamefont {Kroon},\ and\ \citenamefont {Hemminga}}]{van1991conformation}%
  \BibitemOpen
  \bibfield  {author} {\bibinfo {author} {\bibfnamefont {M.}~\bibnamefont {Van~der Graaf}}, \bibinfo {author} {\bibfnamefont {G.~J.}\ \bibnamefont {Kroon}},\ and\ \bibinfo {author} {\bibfnamefont {M.~A.}\ \bibnamefont {Hemminga}},\ }\bibfield  {title} {\bibinfo {title} {Conformation and mobility of the {RNA}-binding {N}-terminal part of the intact coat protein of cowpea chlorotic mottle virus: a two-dimensional proton nuclear magnetic resonance study},\ }\href@noop {} {\bibfield  {journal} {\bibinfo  {journal} {Journal of molecular biology}\ }\textbf {\bibinfo {volume} {220}},\ \bibinfo {pages} {701} (\bibinfo {year} {1991})}\BibitemShut {NoStop}%
\bibitem [{\citenamefont {Nap}\ \emph {et~al.}(2014)\citenamefont {Nap}, \citenamefont {Bo{\v{z}}i{\v{c}}}, \citenamefont {Szleifer},\ and\ \citenamefont {Podgornik}}]{nap2014role}%
  \BibitemOpen
  \bibfield  {author} {\bibinfo {author} {\bibfnamefont {R.~J.}\ \bibnamefont {Nap}}, \bibinfo {author} {\bibfnamefont {A.~L.}\ \bibnamefont {Bo{\v{z}}i{\v{c}}}}, \bibinfo {author} {\bibfnamefont {I.}~\bibnamefont {Szleifer}},\ and\ \bibinfo {author} {\bibfnamefont {R.}~\bibnamefont {Podgornik}},\ }\bibfield  {title} {\bibinfo {title} {The role of solution conditions in the bacteriophage pp7 capsid charge regulation},\ }\href@noop {} {\bibfield  {journal} {\bibinfo  {journal} {Biophysical journal}\ }\textbf {\bibinfo {volume} {107}},\ \bibinfo {pages} {1970} (\bibinfo {year} {2014})}\BibitemShut {NoStop}%
\bibitem [{\citenamefont {Hu}\ \emph {et~al.}(2008{\natexlab{a}})\citenamefont {Hu}, \citenamefont {Zandi}, \citenamefont {Anavitarte}, \citenamefont {Knobler},\ and\ \citenamefont {Gelbart}}]{hu2008packaging}%
  \BibitemOpen
  \bibfield  {author} {\bibinfo {author} {\bibfnamefont {Y.}~\bibnamefont {Hu}}, \bibinfo {author} {\bibfnamefont {R.}~\bibnamefont {Zandi}}, \bibinfo {author} {\bibfnamefont {A.}~\bibnamefont {Anavitarte}}, \bibinfo {author} {\bibfnamefont {C.~M.}\ \bibnamefont {Knobler}},\ and\ \bibinfo {author} {\bibfnamefont {W.~M.}\ \bibnamefont {Gelbart}},\ }\bibfield  {title} {\bibinfo {title} {Packaging of a polymer by a viral capsid: the interplay between polymer length and capsid size},\ }\href@noop {} {\bibfield  {journal} {\bibinfo  {journal} {Biophysical journal}\ }\textbf {\bibinfo {volume} {94}},\ \bibinfo {pages} {1428} (\bibinfo {year} {2008}{\natexlab{a}})}\BibitemShut {NoStop}%
\bibitem [{\citenamefont {Sun}\ \emph {et~al.}(2010)\citenamefont {Sun}, \citenamefont {Rao},\ and\ \citenamefont {Rossmann}}]{sun2010genome}%
  \BibitemOpen
  \bibfield  {author} {\bibinfo {author} {\bibfnamefont {S.}~\bibnamefont {Sun}}, \bibinfo {author} {\bibfnamefont {V.~B.}\ \bibnamefont {Rao}},\ and\ \bibinfo {author} {\bibfnamefont {M.~G.}\ \bibnamefont {Rossmann}},\ }\bibfield  {title} {\bibinfo {title} {Genome packaging in viruses},\ }\href@noop {} {\bibfield  {journal} {\bibinfo  {journal} {Current opinion in structural biology}\ }\textbf {\bibinfo {volume} {20}},\ \bibinfo {pages} {114} (\bibinfo {year} {2010})}\BibitemShut {NoStop}%
\bibitem [{\citenamefont {van~der Schoot}\ and\ \citenamefont {Zandi}(2013)}]{van2013impact}%
  \BibitemOpen
  \bibfield  {author} {\bibinfo {author} {\bibfnamefont {P.}~\bibnamefont {van~der Schoot}}\ and\ \bibinfo {author} {\bibfnamefont {R.}~\bibnamefont {Zandi}},\ }\bibfield  {title} {\bibinfo {title} {Impact of the topology of viral {RNA}s on their encapsulation by virus coat proteins},\ }\href@noop {} {\bibfield  {journal} {\bibinfo  {journal} {Journal of Biological Physics}\ }\textbf {\bibinfo {volume} {39}},\ \bibinfo {pages} {289} (\bibinfo {year} {2013})}\BibitemShut {NoStop}%
\bibitem [{\citenamefont {Kivenson}\ and\ \citenamefont {Hagan}(2010)}]{kivenson2010mechanisms}%
  \BibitemOpen
  \bibfield  {author} {\bibinfo {author} {\bibfnamefont {A.}~\bibnamefont {Kivenson}}\ and\ \bibinfo {author} {\bibfnamefont {M.~F.}\ \bibnamefont {Hagan}},\ }\bibfield  {title} {\bibinfo {title} {Mechanisms of capsid assembly around a polymer},\ }\href@noop {} {\bibfield  {journal} {\bibinfo  {journal} {Biophysical journal}\ }\textbf {\bibinfo {volume} {99}},\ \bibinfo {pages} {619} (\bibinfo {year} {2010})}\BibitemShut {NoStop}%
\bibitem [{\citenamefont {Garmann}\ \emph {et~al.}(2014)\citenamefont {Garmann}, \citenamefont {Comas-Garcia}, \citenamefont {Koay}, \citenamefont {Cornelissen}, \citenamefont {Knobler},\ and\ \citenamefont {Gelbart}}]{garmann2014role}%
  \BibitemOpen
  \bibfield  {author} {\bibinfo {author} {\bibfnamefont {R.~F.}\ \bibnamefont {Garmann}}, \bibinfo {author} {\bibfnamefont {M.}~\bibnamefont {Comas-Garcia}}, \bibinfo {author} {\bibfnamefont {M.~S.}\ \bibnamefont {Koay}}, \bibinfo {author} {\bibfnamefont {J.~J.}\ \bibnamefont {Cornelissen}}, \bibinfo {author} {\bibfnamefont {C.~M.}\ \bibnamefont {Knobler}},\ and\ \bibinfo {author} {\bibfnamefont {W.~M.}\ \bibnamefont {Gelbart}},\ }\bibfield  {title} {\bibinfo {title} {Role of electrostatics in the assembly pathway of a single-stranded {RNA} virus},\ }\href@noop {} {\bibfield  {journal} {\bibinfo  {journal} {Journal of Virology}\ }\textbf {\bibinfo {volume} {88}},\ \bibinfo {pages} {10472} (\bibinfo {year} {2014})}\BibitemShut {NoStop}%
\bibitem [{\citenamefont {Erdemci-Tandogan}\ \emph {et~al.}(2014)\citenamefont {Erdemci-Tandogan}, \citenamefont {Wagner}, \citenamefont {van~der Schoot}, \citenamefont {Podgornik},\ and\ \citenamefont {Zandi}}]{erdemci2014rna}%
  \BibitemOpen
  \bibfield  {author} {\bibinfo {author} {\bibfnamefont {G.}~\bibnamefont {Erdemci-Tandogan}}, \bibinfo {author} {\bibfnamefont {J.}~\bibnamefont {Wagner}}, \bibinfo {author} {\bibfnamefont {P.}~\bibnamefont {van~der Schoot}}, \bibinfo {author} {\bibfnamefont {R.}~\bibnamefont {Podgornik}},\ and\ \bibinfo {author} {\bibfnamefont {R.}~\bibnamefont {Zandi}},\ }\bibfield  {title} {\bibinfo {title} {{RNA} topology remolds electrostatic stabilization of viruses},\ }\href@noop {} {\bibfield  {journal} {\bibinfo  {journal} {Physical Review E}\ }\textbf {\bibinfo {volume} {89}},\ \bibinfo {pages} {032707} (\bibinfo {year} {2014})}\BibitemShut {NoStop}%
\bibitem [{\citenamefont {Erdemci-Tandogan}\ \emph {et~al.}(2016{\natexlab{a}})\citenamefont {Erdemci-Tandogan}, \citenamefont {Wagner}, \citenamefont {van~der Schoot}, \citenamefont {Podgornik},\ and\ \citenamefont {Zandi}}]{erdemci2016effects}%
  \BibitemOpen
  \bibfield  {author} {\bibinfo {author} {\bibfnamefont {G.}~\bibnamefont {Erdemci-Tandogan}}, \bibinfo {author} {\bibfnamefont {J.}~\bibnamefont {Wagner}}, \bibinfo {author} {\bibfnamefont {P.}~\bibnamefont {van~der Schoot}}, \bibinfo {author} {\bibfnamefont {R.}~\bibnamefont {Podgornik}},\ and\ \bibinfo {author} {\bibfnamefont {R.}~\bibnamefont {Zandi}},\ }\bibfield  {title} {\bibinfo {title} {Effects of {RNA} branching on the electrostatic stabilization of viruses},\ }\href@noop {} {\bibfield  {journal} {\bibinfo  {journal} {Physical Review E}\ }\textbf {\bibinfo {volume} {94}},\ \bibinfo {pages} {022408} (\bibinfo {year} {2016}{\natexlab{a}})}\BibitemShut {NoStop}%
\bibitem [{\citenamefont {French}\ \emph {et~al.}(2010)\citenamefont {French}, \citenamefont {Parsegian}, \citenamefont {Podgornik}, \citenamefont {Rajter}, \citenamefont {Jagota}, \citenamefont {Luo}, \citenamefont {Asthagiri}, \citenamefont {Chaudhury}, \citenamefont {Chiang}, \citenamefont {Granick} \emph {et~al.}}]{french2010long}%
  \BibitemOpen
  \bibfield  {author} {\bibinfo {author} {\bibfnamefont {R.~H.}\ \bibnamefont {French}}, \bibinfo {author} {\bibfnamefont {V.~A.}\ \bibnamefont {Parsegian}}, \bibinfo {author} {\bibfnamefont {R.}~\bibnamefont {Podgornik}}, \bibinfo {author} {\bibfnamefont {R.~F.}\ \bibnamefont {Rajter}}, \bibinfo {author} {\bibfnamefont {A.}~\bibnamefont {Jagota}}, \bibinfo {author} {\bibfnamefont {J.}~\bibnamefont {Luo}}, \bibinfo {author} {\bibfnamefont {D.}~\bibnamefont {Asthagiri}}, \bibinfo {author} {\bibfnamefont {M.~K.}\ \bibnamefont {Chaudhury}}, \bibinfo {author} {\bibfnamefont {Y.-m.}\ \bibnamefont {Chiang}}, \bibinfo {author} {\bibfnamefont {S.}~\bibnamefont {Granick}}, \emph {et~al.},\ }\bibfield  {title} {\bibinfo {title} {Long range interactions in nanoscale science},\ }\href@noop {} {\bibfield  {journal} {\bibinfo  {journal} {Reviews of Modern Physics}\ }\textbf {\bibinfo {volume} {82}},\ \bibinfo {pages} {1887} (\bibinfo {year} {2010})}\BibitemShut {NoStop}%
\bibitem [{\citenamefont {Bo{\v{z}}i{\v{c}}}\ and\ \citenamefont {Podgornik}(2018)}]{bovzivc2017varieties}%
  \BibitemOpen
  \bibfield  {author} {\bibinfo {author} {\bibfnamefont {A.~L.}\ \bibnamefont {Bo{\v{z}}i{\v{c}}}}\ and\ \bibinfo {author} {\bibfnamefont {R.}~\bibnamefont {Podgornik}},\ }\bibfield  {title} {\bibinfo {title} {Varieties of charge distributions in coat proteins of ss{RNA}+ viruses},\ }\href@noop {} {\bibfield  {journal} {\bibinfo  {journal} {Journal of Physics: Condensed Matter}\ }\textbf {\bibinfo {volume} {30}},\ \bibinfo {pages} {024001} (\bibinfo {year} {2018})}\BibitemShut {NoStop}%
\bibitem [{\citenamefont {Douglas}\ and\ \citenamefont {Young}(1998)}]{douglas1998host}%
  \BibitemOpen
  \bibfield  {author} {\bibinfo {author} {\bibfnamefont {T.}~\bibnamefont {Douglas}}\ and\ \bibinfo {author} {\bibfnamefont {M.}~\bibnamefont {Young}},\ }\bibfield  {title} {\bibinfo {title} {Host--guest encapsulation of materials by assembled virus protein cages},\ }\href@noop {} {\bibfield  {journal} {\bibinfo  {journal} {Nature}\ }\textbf {\bibinfo {volume} {393}},\ \bibinfo {pages} {152} (\bibinfo {year} {1998})}\BibitemShut {NoStop}%
\bibitem [{\citenamefont {Maassen}\ \emph {et~al.}(2019)\citenamefont {Maassen}, \citenamefont {Huskens},\ and\ \citenamefont {Cornelissen}}]{maassen2019elucidating}%
  \BibitemOpen
  \bibfield  {author} {\bibinfo {author} {\bibfnamefont {S.~J.}\ \bibnamefont {Maassen}}, \bibinfo {author} {\bibfnamefont {J.}~\bibnamefont {Huskens}},\ and\ \bibinfo {author} {\bibfnamefont {J.~J.}\ \bibnamefont {Cornelissen}},\ }\bibfield  {title} {\bibinfo {title} {Elucidating the thermodynamic driving forces of polyanion-templated virus-like particle assembly},\ }\href@noop {} {\bibfield  {journal} {\bibinfo  {journal} {The journal of physical chemistry B}\ }\textbf {\bibinfo {volume} {123}},\ \bibinfo {pages} {9733} (\bibinfo {year} {2019})}\BibitemShut {NoStop}%
\bibitem [{\citenamefont {Wang}\ \emph {et~al.}(2006)\citenamefont {Wang}, \citenamefont {He}, \citenamefont {Tong}, \citenamefont {Liu}, \citenamefont {Ren},\ and\ \citenamefont {Zeng}}]{wang2006combination}%
  \BibitemOpen
  \bibfield  {author} {\bibinfo {author} {\bibfnamefont {C.}~\bibnamefont {Wang}}, \bibinfo {author} {\bibfnamefont {C.}~\bibnamefont {He}}, \bibinfo {author} {\bibfnamefont {Z.}~\bibnamefont {Tong}}, \bibinfo {author} {\bibfnamefont {X.}~\bibnamefont {Liu}}, \bibinfo {author} {\bibfnamefont {B.}~\bibnamefont {Ren}},\ and\ \bibinfo {author} {\bibfnamefont {F.}~\bibnamefont {Zeng}},\ }\bibfield  {title} {\bibinfo {title} {Combination of adsorption by porous {CaCO3} microparticles and encapsulation by polyelectrolyte multilayer films for sustained drug delivery},\ }\href@noop {} {\bibfield  {journal} {\bibinfo  {journal} {International journal of pharmaceutics}\ }\textbf {\bibinfo {volume} {308}},\ \bibinfo {pages} {160} (\bibinfo {year} {2006})}\BibitemShut {NoStop}%
\bibitem [{\citenamefont {Chang}\ \emph {et~al.}(2008)\citenamefont {Chang}, \citenamefont {Knobler}, \citenamefont {Gelbart},\ and\ \citenamefont {Mason}}]{chang2008curvature}%
  \BibitemOpen
  \bibfield  {author} {\bibinfo {author} {\bibfnamefont {C.~B.}\ \bibnamefont {Chang}}, \bibinfo {author} {\bibfnamefont {C.~M.}\ \bibnamefont {Knobler}}, \bibinfo {author} {\bibfnamefont {W.~M.}\ \bibnamefont {Gelbart}},\ and\ \bibinfo {author} {\bibfnamefont {T.~G.}\ \bibnamefont {Mason}},\ }\bibfield  {title} {\bibinfo {title} {Curvature dependence of viral protein structures on encapsidated nanoemulsion droplets},\ }\href@noop {} {\bibfield  {journal} {\bibinfo  {journal} {Acs Nano}\ }\textbf {\bibinfo {volume} {2}},\ \bibinfo {pages} {281} (\bibinfo {year} {2008})}\BibitemShut {NoStop}%
\bibitem [{\citenamefont {Lin}\ \emph {et~al.}(2012)\citenamefont {Lin}, \citenamefont {Van Der~Schoot},\ and\ \citenamefont {Zandi}}]{Hsiang-Ku}%
  \BibitemOpen
  \bibfield  {author} {\bibinfo {author} {\bibfnamefont {H.-K.}\ \bibnamefont {Lin}}, \bibinfo {author} {\bibfnamefont {P.}~\bibnamefont {Van Der~Schoot}},\ and\ \bibinfo {author} {\bibfnamefont {R.}~\bibnamefont {Zandi}},\ }\bibfield  {title} {\bibinfo {title} {{Impact of Charge Variation on the Encapsulation of Nanoparticles by Virus Coat Proteins}},\ }\href {https://doi.org/10.1088/1478-3975/9/6/066004} {\bibfield  {journal} {\bibinfo  {journal} {Phys. Biol.}\ }\textbf {\bibinfo {volume} {9}},\ \bibinfo {pages} {066004} (\bibinfo {year} {2012})}\BibitemShut {NoStop}%
\bibitem [{\citenamefont {Perlmutter}\ \emph {et~al.}(2013)\citenamefont {Perlmutter}, \citenamefont {Qiao},\ and\ \citenamefont {Hagan}}]{perlmutter2013viral}%
  \BibitemOpen
  \bibfield  {author} {\bibinfo {author} {\bibfnamefont {J.~D.}\ \bibnamefont {Perlmutter}}, \bibinfo {author} {\bibfnamefont {C.}~\bibnamefont {Qiao}},\ and\ \bibinfo {author} {\bibfnamefont {M.~F.}\ \bibnamefont {Hagan}},\ }\bibfield  {title} {\bibinfo {title} {Viral genome structures are optimal for capsid assembly},\ }\href@noop {} {\bibfield  {journal} {\bibinfo  {journal} {elife}\ }\textbf {\bibinfo {volume} {2}},\ \bibinfo {pages} {e00632} (\bibinfo {year} {2013})}\BibitemShut {NoStop}%
\bibitem [{\citenamefont {Perlmutter}\ and\ \citenamefont {Hagan}(2015{\natexlab{b}})}]{perlmutter2015mechanisms}%
  \BibitemOpen
  \bibfield  {author} {\bibinfo {author} {\bibfnamefont {J.~D.}\ \bibnamefont {Perlmutter}}\ and\ \bibinfo {author} {\bibfnamefont {M.~F.}\ \bibnamefont {Hagan}},\ }\bibfield  {title} {\bibinfo {title} {Mechanisms of virus assembly},\ }\href@noop {} {\bibfield  {journal} {\bibinfo  {journal} {Annual review of physical chemistry}\ }\textbf {\bibinfo {volume} {66}},\ \bibinfo {pages} {217} (\bibinfo {year} {2015}{\natexlab{b}})}\BibitemShut {NoStop}%
\bibitem [{\citenamefont {Kundagrami}\ and\ \citenamefont {Muthukumar}(2008)}]{kundagrami2008theory}%
  \BibitemOpen
  \bibfield  {author} {\bibinfo {author} {\bibfnamefont {A.}~\bibnamefont {Kundagrami}}\ and\ \bibinfo {author} {\bibfnamefont {M.}~\bibnamefont {Muthukumar}},\ }\bibfield  {title} {\bibinfo {title} {Theory of competitive counterion adsorption on flexible polyelectrolytes: Divalent salts},\ }\href@noop {} {\bibfield  {journal} {\bibinfo  {journal} {The Journal of chemical physics}\ }\textbf {\bibinfo {volume} {128}} (\bibinfo {year} {2008})}\BibitemShut {NoStop}%
\bibitem [{\citenamefont {Muthukumar}(2017)}]{muthukumar201750th}%
  \BibitemOpen
  \bibfield  {author} {\bibinfo {author} {\bibfnamefont {M.}~\bibnamefont {Muthukumar}},\ }\bibfield  {title} {\bibinfo {title} {50th anniversary perspective: A perspective on polyelectrolyte solutions},\ }\href@noop {} {\bibfield  {journal} {\bibinfo  {journal} {Macromolecules}\ }\textbf {\bibinfo {volume} {50}},\ \bibinfo {pages} {9528} (\bibinfo {year} {2017})}\BibitemShut {NoStop}%
\bibitem [{\citenamefont {Ting}\ \emph {et~al.}(2011)\citenamefont {Ting}, \citenamefont {Wu},\ and\ \citenamefont {Wang}}]{ting2011thermodynamic}%
  \BibitemOpen
  \bibfield  {author} {\bibinfo {author} {\bibfnamefont {C.~L.}\ \bibnamefont {Ting}}, \bibinfo {author} {\bibfnamefont {J.}~\bibnamefont {Wu}},\ and\ \bibinfo {author} {\bibfnamefont {Z.-G.}\ \bibnamefont {Wang}},\ }\bibfield  {title} {\bibinfo {title} {Thermodynamic basis for the genome to capsid charge relationship in viral encapsidation},\ }\href@noop {} {\bibfield  {journal} {\bibinfo  {journal} {Proceedings of the National Academy of Sciences}\ }\textbf {\bibinfo {volume} {108}},\ \bibinfo {pages} {16986} (\bibinfo {year} {2011})}\BibitemShut {NoStop}%
\bibitem [{\citenamefont {Hu}\ \emph {et~al.}(2008{\natexlab{b}})\citenamefont {Hu}, \citenamefont {Zhang},\ and\ \citenamefont {Shklovskii}}]{hu2008electrostatic}%
  \BibitemOpen
  \bibfield  {author} {\bibinfo {author} {\bibfnamefont {T.}~\bibnamefont {Hu}}, \bibinfo {author} {\bibfnamefont {R.}~\bibnamefont {Zhang}},\ and\ \bibinfo {author} {\bibfnamefont {B.}~\bibnamefont {Shklovskii}},\ }\bibfield  {title} {\bibinfo {title} {Electrostatic theory of viral self-assembly},\ }\href@noop {} {\bibfield  {journal} {\bibinfo  {journal} {Physica A: Statistical Mechanics and its Applications}\ }\textbf {\bibinfo {volume} {387}},\ \bibinfo {pages} {3059} (\bibinfo {year} {2008}{\natexlab{b}})}\BibitemShut {NoStop}%
\bibitem [{\citenamefont {Muhren}\ and\ \citenamefont {van~der Schoot}(2023)}]{muhren2023electrostatic}%
  \BibitemOpen
  \bibfield  {author} {\bibinfo {author} {\bibfnamefont {H.}~\bibnamefont {Muhren}}\ and\ \bibinfo {author} {\bibfnamefont {P.}~\bibnamefont {van~der Schoot}},\ }\bibfield  {title} {\bibinfo {title} {Electrostatic theory of the acidity of the solution in the lumina of viruses and virus-like particles},\ }\href@noop {} {\bibfield  {journal} {\bibinfo  {journal} {The Journal of Physical Chemistry B}\ }\textbf {\bibinfo {volume} {127}},\ \bibinfo {pages} {2160} (\bibinfo {year} {2023})}\BibitemShut {NoStop}%
\bibitem [{\citenamefont {Bruinsma}\ \emph {et~al.}(2016)\citenamefont {Bruinsma}, \citenamefont {Comas-Garcia}, \citenamefont {Garmann},\ and\ \citenamefont {Grosberg}}]{bruinsma2016equilibrium}%
  \BibitemOpen
  \bibfield  {author} {\bibinfo {author} {\bibfnamefont {R.~F.}\ \bibnamefont {Bruinsma}}, \bibinfo {author} {\bibfnamefont {M.}~\bibnamefont {Comas-Garcia}}, \bibinfo {author} {\bibfnamefont {R.~F.}\ \bibnamefont {Garmann}},\ and\ \bibinfo {author} {\bibfnamefont {A.~Y.}\ \bibnamefont {Grosberg}},\ }\bibfield  {title} {\bibinfo {title} {Equilibrium self-assembly of small {RNA} viruses},\ }\href@noop {} {\bibfield  {journal} {\bibinfo  {journal} {Physical Review E}\ }\textbf {\bibinfo {volume} {93}},\ \bibinfo {pages} {032405} (\bibinfo {year} {2016})}\BibitemShut {NoStop}%
\bibitem [{\citenamefont {Joanny}(1999)}]{joanny1999polyelectrolyte}%
  \BibitemOpen
  \bibfield  {author} {\bibinfo {author} {\bibfnamefont {J.}~\bibnamefont {Joanny}},\ }\bibfield  {title} {\bibinfo {title} {Polyelectrolyte adsorption and charge inversion},\ }\href@noop {} {\bibfield  {journal} {\bibinfo  {journal} {The European Physical Journal B-Condensed Matter and Complex Systems}\ }\textbf {\bibinfo {volume} {9}},\ \bibinfo {pages} {117} (\bibinfo {year} {1999})}\BibitemShut {NoStop}%
\bibitem [{\citenamefont {Van~der Schee}\ and\ \citenamefont {Lyklema}(1984)}]{van1984lattice}%
  \BibitemOpen
  \bibfield  {author} {\bibinfo {author} {\bibfnamefont {H.}~\bibnamefont {Van~der Schee}}\ and\ \bibinfo {author} {\bibfnamefont {J.}~\bibnamefont {Lyklema}},\ }\bibfield  {title} {\bibinfo {title} {A lattice theory of polyelectrolyte adsorption},\ }\href@noop {} {\bibfield  {journal} {\bibinfo  {journal} {The Journal of Physical Chemistry}\ }\textbf {\bibinfo {volume} {88}},\ \bibinfo {pages} {6661} (\bibinfo {year} {1984})}\BibitemShut {NoStop}%
\bibitem [{\citenamefont {Szilagyi}\ \emph {et~al.}(2014)\citenamefont {Szilagyi}, \citenamefont {Trefalt}, \citenamefont {Tiraferri}, \citenamefont {Maroni},\ and\ \citenamefont {Borkovec}}]{szilagyi2014polyelectrolyte}%
  \BibitemOpen
  \bibfield  {author} {\bibinfo {author} {\bibfnamefont {I.}~\bibnamefont {Szilagyi}}, \bibinfo {author} {\bibfnamefont {G.}~\bibnamefont {Trefalt}}, \bibinfo {author} {\bibfnamefont {A.}~\bibnamefont {Tiraferri}}, \bibinfo {author} {\bibfnamefont {P.}~\bibnamefont {Maroni}},\ and\ \bibinfo {author} {\bibfnamefont {M.}~\bibnamefont {Borkovec}},\ }\bibfield  {title} {\bibinfo {title} {Polyelectrolyte adsorption, interparticle forces, and colloidal aggregation},\ }\href@noop {} {\bibfield  {journal} {\bibinfo  {journal} {Soft Matter}\ }\textbf {\bibinfo {volume} {10}},\ \bibinfo {pages} {2479} (\bibinfo {year} {2014})}\BibitemShut {NoStop}%
\bibitem [{\citenamefont {Fingerhut}\ \emph {et~al.}(2021)\citenamefont {Fingerhut}, \citenamefont {Schauss}, \citenamefont {Kundu},\ and\ \citenamefont {Elsaesser}}]{fingerhut2021contact}%
  \BibitemOpen
  \bibfield  {author} {\bibinfo {author} {\bibfnamefont {B.~P.}\ \bibnamefont {Fingerhut}}, \bibinfo {author} {\bibfnamefont {J.}~\bibnamefont {Schauss}}, \bibinfo {author} {\bibfnamefont {A.}~\bibnamefont {Kundu}},\ and\ \bibinfo {author} {\bibfnamefont {T.}~\bibnamefont {Elsaesser}},\ }\bibfield  {title} {\bibinfo {title} {Contact pairs of {RNA} with magnesium ions-electrostatics beyond the poisson-boltzmann equation},\ }\href@noop {} {\bibfield  {journal} {\bibinfo  {journal} {Biophysical Journal}\ }\textbf {\bibinfo {volume} {120}},\ \bibinfo {pages} {5322} (\bibinfo {year} {2021})}\BibitemShut {NoStop}%
\bibitem [{\citenamefont {Kirmizialtin}\ \emph {et~al.}(2012)\citenamefont {Kirmizialtin}, \citenamefont {Silalahi}, \citenamefont {Elber},\ and\ \citenamefont {Fenley}}]{kirmizialtin2012ionic}%
  \BibitemOpen
  \bibfield  {author} {\bibinfo {author} {\bibfnamefont {S.}~\bibnamefont {Kirmizialtin}}, \bibinfo {author} {\bibfnamefont {A.~R.}\ \bibnamefont {Silalahi}}, \bibinfo {author} {\bibfnamefont {R.}~\bibnamefont {Elber}},\ and\ \bibinfo {author} {\bibfnamefont {M.~O.}\ \bibnamefont {Fenley}},\ }\bibfield  {title} {\bibinfo {title} {The ionic atmosphere around {A-RNA}: Poisson-boltzmann and molecular dynamics simulations},\ }\href@noop {} {\bibfield  {journal} {\bibinfo  {journal} {Biophysical journal}\ }\textbf {\bibinfo {volume} {102}},\ \bibinfo {pages} {829} (\bibinfo {year} {2012})}\BibitemShut {NoStop}%
\bibitem [{\citenamefont {Nguyen}\ \emph {et~al.}(2017)\citenamefont {Nguyen}, \citenamefont {Wang},\ and\ \citenamefont {Wei}}]{nguyen2017accurate}%
  \BibitemOpen
  \bibfield  {author} {\bibinfo {author} {\bibfnamefont {D.~D.}\ \bibnamefont {Nguyen}}, \bibinfo {author} {\bibfnamefont {B.}~\bibnamefont {Wang}},\ and\ \bibinfo {author} {\bibfnamefont {G.-W.}\ \bibnamefont {Wei}},\ }\bibfield  {title} {\bibinfo {title} {Accurate, robust, and reliable calculations of poisson--boltzmann binding energies},\ }\href@noop {} {\bibfield  {journal} {\bibinfo  {journal} {Journal of computational chemistry}\ }\textbf {\bibinfo {volume} {38}},\ \bibinfo {pages} {941} (\bibinfo {year} {2017})}\BibitemShut {NoStop}%
\bibitem [{\citenamefont {Netz}(2001)}]{netz2001electrostatistics}%
  \BibitemOpen
  \bibfield  {author} {\bibinfo {author} {\bibfnamefont {R.~R.}\ \bibnamefont {Netz}},\ }\bibfield  {title} {\bibinfo {title} {Electrostatistics of counter-ions at and between planar charged walls: From poisson-boltzmann to the strong-coupling theory},\ }\href@noop {} {\bibfield  {journal} {\bibinfo  {journal} {The European Physical Journal E}\ }\textbf {\bibinfo {volume} {5}},\ \bibinfo {pages} {557} (\bibinfo {year} {2001})}\BibitemShut {NoStop}%
\bibitem [{\citenamefont {Kronenberg}\ \emph {et~al.}(2001)\citenamefont {Kronenberg}, \citenamefont {Kleinschmidt},\ and\ \citenamefont {B{\"o}ttcher}}]{kronenberg2001electron}%
  \BibitemOpen
  \bibfield  {author} {\bibinfo {author} {\bibfnamefont {S.}~\bibnamefont {Kronenberg}}, \bibinfo {author} {\bibfnamefont {J.~A.}\ \bibnamefont {Kleinschmidt}},\ and\ \bibinfo {author} {\bibfnamefont {B.}~\bibnamefont {B{\"o}ttcher}},\ }\bibfield  {title} {\bibinfo {title} {Electron cryo-microscopy and image reconstruction of adeno-associated virus type 2 empty capsids},\ }\href@noop {} {\bibfield  {journal} {\bibinfo  {journal} {EMBO reports}\ }\textbf {\bibinfo {volume} {2}},\ \bibinfo {pages} {997} (\bibinfo {year} {2001})}\BibitemShut {NoStop}%
\bibitem [{\citenamefont {Li}\ \emph {et~al.}(2017)\citenamefont {Li}, \citenamefont {Erdemci-Tandogan}, \citenamefont {Wagner}, \citenamefont {van~der Schoot},\ and\ \citenamefont {Zandi}}]{li2017impact}%
  \BibitemOpen
  \bibfield  {author} {\bibinfo {author} {\bibfnamefont {S.}~\bibnamefont {Li}}, \bibinfo {author} {\bibfnamefont {G.}~\bibnamefont {Erdemci-Tandogan}}, \bibinfo {author} {\bibfnamefont {J.}~\bibnamefont {Wagner}}, \bibinfo {author} {\bibfnamefont {P.}~\bibnamefont {van~der Schoot}},\ and\ \bibinfo {author} {\bibfnamefont {R.}~\bibnamefont {Zandi}},\ }\bibfield  {title} {\bibinfo {title} {Impact of a nonuniform charge distribution on virus assembly},\ }\href@noop {} {\bibfield  {journal} {\bibinfo  {journal} {Physical Review E}\ }\textbf {\bibinfo {volume} {96}},\ \bibinfo {pages} {022401} (\bibinfo {year} {2017})}\BibitemShut {NoStop}%
\bibitem [{\citenamefont {Forrey}\ and\ \citenamefont {Muthukumar}(2009)}]{forrey2009electrostatics}%
  \BibitemOpen
  \bibfield  {author} {\bibinfo {author} {\bibfnamefont {C.}~\bibnamefont {Forrey}}\ and\ \bibinfo {author} {\bibfnamefont {M.}~\bibnamefont {Muthukumar}},\ }\bibfield  {title} {\bibinfo {title} {Electrostatics of capsid-induced viral {RNA} organization},\ }\href@noop {} {\bibfield  {journal} {\bibinfo  {journal} {The Journal of chemical physics}\ }\textbf {\bibinfo {volume} {131}} (\bibinfo {year} {2009})}\BibitemShut {NoStop}%
\bibitem [{\citenamefont {Wu}\ \emph {et~al.}(2003)\citenamefont {Wu}, \citenamefont {Smith}, \citenamefont {Wieczorek}, \citenamefont {Mather}, \citenamefont {Duncan}, \citenamefont {White}, \citenamefont {McGill}, \citenamefont {Katten},\ and\ \citenamefont {Heller}}]{wu2003biological}%
  \BibitemOpen
  \bibfield  {author} {\bibinfo {author} {\bibfnamefont {A.~H.}\ \bibnamefont {Wu}}, \bibinfo {author} {\bibfnamefont {A.}~\bibnamefont {Smith}}, \bibinfo {author} {\bibfnamefont {S.}~\bibnamefont {Wieczorek}}, \bibinfo {author} {\bibfnamefont {J.~F.}\ \bibnamefont {Mather}}, \bibinfo {author} {\bibfnamefont {B.}~\bibnamefont {Duncan}}, \bibinfo {author} {\bibfnamefont {C.~M.}\ \bibnamefont {White}}, \bibinfo {author} {\bibfnamefont {C.}~\bibnamefont {McGill}}, \bibinfo {author} {\bibfnamefont {D.}~\bibnamefont {Katten}},\ and\ \bibinfo {author} {\bibfnamefont {G.}~\bibnamefont {Heller}},\ }\bibfield  {title} {\bibinfo {title} {Biological variation for {N}-terminal pro-and {B-type} natriuretic peptides and implications for therapeutic monitoring of patients with congestive heart failure},\ }\href@noop {} {\bibfield  {journal} {\bibinfo  {journal} {The American journal of cardiology}\ }\textbf {\bibinfo {volume} {92}},\ \bibinfo {pages} {628} (\bibinfo {year} {2003})}\BibitemShut {NoStop}%
\bibitem [{\citenamefont {Adachi}\ \emph {et~al.}(2019)\citenamefont {Adachi}, \citenamefont {Contreras}, \citenamefont {Harant}, \citenamefont {Wu}, \citenamefont {Derevnina}, \citenamefont {Sakai}, \citenamefont {Duggan}, \citenamefont {Moratto}, \citenamefont {Bozkurt}, \citenamefont {Maqbool} \emph {et~al.}}]{adachi2019n}%
  \BibitemOpen
  \bibfield  {author} {\bibinfo {author} {\bibfnamefont {H.}~\bibnamefont {Adachi}}, \bibinfo {author} {\bibfnamefont {M.~P.}\ \bibnamefont {Contreras}}, \bibinfo {author} {\bibfnamefont {A.}~\bibnamefont {Harant}}, \bibinfo {author} {\bibfnamefont {C.-h.}\ \bibnamefont {Wu}}, \bibinfo {author} {\bibfnamefont {L.}~\bibnamefont {Derevnina}}, \bibinfo {author} {\bibfnamefont {T.}~\bibnamefont {Sakai}}, \bibinfo {author} {\bibfnamefont {C.}~\bibnamefont {Duggan}}, \bibinfo {author} {\bibfnamefont {E.}~\bibnamefont {Moratto}}, \bibinfo {author} {\bibfnamefont {T.~O.}\ \bibnamefont {Bozkurt}}, \bibinfo {author} {\bibfnamefont {A.}~\bibnamefont {Maqbool}}, \emph {et~al.},\ }\bibfield  {title} {\bibinfo {title} {An {N}-terminal motif in nlr immune receptors is functionally conserved across distantly related plant species},\ }\href@noop {} {\bibfield  {journal} {\bibinfo  {journal} {Elife}\ }\textbf {\bibinfo {volume} {8}},\ \bibinfo {pages} {e49956} (\bibinfo {year} {2019})}\BibitemShut {NoStop}%
\bibitem [{\citenamefont {Wu}\ \emph {et~al.}(1995)\citenamefont {Wu}, \citenamefont {Knudsen}, \citenamefont {Feller}, \citenamefont {Zheng}, \citenamefont {Sali}, \citenamefont {Cowburn}, \citenamefont {Hanafusa},\ and\ \citenamefont {Kuriyan}}]{wu1995structural}%
  \BibitemOpen
  \bibfield  {author} {\bibinfo {author} {\bibfnamefont {X.}~\bibnamefont {Wu}}, \bibinfo {author} {\bibfnamefont {B.}~\bibnamefont {Knudsen}}, \bibinfo {author} {\bibfnamefont {S.~M.}\ \bibnamefont {Feller}}, \bibinfo {author} {\bibfnamefont {J.}~\bibnamefont {Zheng}}, \bibinfo {author} {\bibfnamefont {A.}~\bibnamefont {Sali}}, \bibinfo {author} {\bibfnamefont {D.}~\bibnamefont {Cowburn}}, \bibinfo {author} {\bibfnamefont {H.}~\bibnamefont {Hanafusa}},\ and\ \bibinfo {author} {\bibfnamefont {J.}~\bibnamefont {Kuriyan}},\ }\bibfield  {title} {\bibinfo {title} {Structural basis for the specific interaction of lysine-containing proline-rich peptides with the {N}-terminal {SH3} domain of {c-Crk}},\ }\href@noop {} {\bibfield  {journal} {\bibinfo  {journal} {Structure}\ }\textbf {\bibinfo {volume} {3}},\ \bibinfo {pages} {215} (\bibinfo {year} {1995})}\BibitemShut {NoStop}%
\bibitem [{\citenamefont {Ni}\ \emph {et~al.}(2012)\citenamefont {Ni}, \citenamefont {Wang}, \citenamefont {Ma}, \citenamefont {Das}, \citenamefont {Sokol}, \citenamefont {Chiu}, \citenamefont {Dragnea}, \citenamefont {Hagan},\ and\ \citenamefont {Kao}}]{ni2012examination}%
  \BibitemOpen
  \bibfield  {author} {\bibinfo {author} {\bibfnamefont {P.}~\bibnamefont {Ni}}, \bibinfo {author} {\bibfnamefont {Z.}~\bibnamefont {Wang}}, \bibinfo {author} {\bibfnamefont {X.}~\bibnamefont {Ma}}, \bibinfo {author} {\bibfnamefont {N.~C.}\ \bibnamefont {Das}}, \bibinfo {author} {\bibfnamefont {P.}~\bibnamefont {Sokol}}, \bibinfo {author} {\bibfnamefont {W.}~\bibnamefont {Chiu}}, \bibinfo {author} {\bibfnamefont {B.}~\bibnamefont {Dragnea}}, \bibinfo {author} {\bibfnamefont {M.}~\bibnamefont {Hagan}},\ and\ \bibinfo {author} {\bibfnamefont {C.~C.}\ \bibnamefont {Kao}},\ }\bibfield  {title} {\bibinfo {title} {An examination of the electrostatic interactions between the {N}-terminal tail of the brome mosaic virus coat protein and encapsidated {RNAs}},\ }\href@noop {} {\bibfield  {journal} {\bibinfo  {journal} {Journal of molecular biology}\ }\textbf {\bibinfo {volume} {419}},\ \bibinfo {pages} {284} (\bibinfo {year} {2012})}\BibitemShut {NoStop}%
\bibitem [{\citenamefont {Dong}\ \emph {et~al.}(2020)\citenamefont {Dong}, \citenamefont {Li},\ and\ \citenamefont {Zandi}}]{dong2020effect}%
  \BibitemOpen
  \bibfield  {author} {\bibinfo {author} {\bibfnamefont {Y.}~\bibnamefont {Dong}}, \bibinfo {author} {\bibfnamefont {S.}~\bibnamefont {Li}},\ and\ \bibinfo {author} {\bibfnamefont {R.}~\bibnamefont {Zandi}},\ }\bibfield  {title} {\bibinfo {title} {Effect of the charge distribution of virus coat proteins on the length of packaged {RNAs}},\ }\href@noop {} {\bibfield  {journal} {\bibinfo  {journal} {Physical Review E}\ }\textbf {\bibinfo {volume} {102}},\ \bibinfo {pages} {062423} (\bibinfo {year} {2020})}\BibitemShut {NoStop}%
\bibitem [{\citenamefont {Schwer}\ \emph {et~al.}(2004)\citenamefont {Schwer}, \citenamefont {Ren}, \citenamefont {Pietschmann}, \citenamefont {Kartenbeck}, \citenamefont {Kaehlcke}, \citenamefont {Bartenschlager}, \citenamefont {Yen},\ and\ \citenamefont {Ott}}]{schwer2004targeting}%
  \BibitemOpen
  \bibfield  {author} {\bibinfo {author} {\bibfnamefont {B.}~\bibnamefont {Schwer}}, \bibinfo {author} {\bibfnamefont {S.}~\bibnamefont {Ren}}, \bibinfo {author} {\bibfnamefont {T.}~\bibnamefont {Pietschmann}}, \bibinfo {author} {\bibfnamefont {J.}~\bibnamefont {Kartenbeck}}, \bibinfo {author} {\bibfnamefont {K.}~\bibnamefont {Kaehlcke}}, \bibinfo {author} {\bibfnamefont {R.}~\bibnamefont {Bartenschlager}}, \bibinfo {author} {\bibfnamefont {T.~B.}\ \bibnamefont {Yen}},\ and\ \bibinfo {author} {\bibfnamefont {M.}~\bibnamefont {Ott}},\ }\bibfield  {title} {\bibinfo {title} {Targeting of hepatitis {C} virus core protein to mitochondria through a novel {C-terminal} localization motif},\ }\href@noop {} {\bibfield  {journal} {\bibinfo  {journal} {Journal of virology}\ }\textbf {\bibinfo {volume} {78}},\ \bibinfo {pages} {7958} (\bibinfo {year} {2004})}\BibitemShut {NoStop}%
\bibitem [{\citenamefont {Tresset}(2009)}]{tresset2009multiple}%
  \BibitemOpen
  \bibfield  {author} {\bibinfo {author} {\bibfnamefont {G.}~\bibnamefont {Tresset}},\ }\bibfield  {title} {\bibinfo {title} {The multiple faces of self-assembled lipidic systems},\ }\href@noop {} {\bibfield  {journal} {\bibinfo  {journal} {PMC biophysics}\ }\textbf {\bibinfo {volume} {2}},\ \bibinfo {pages} {1} (\bibinfo {year} {2009})}\BibitemShut {NoStop}%
\bibitem [{\citenamefont {Knobler}\ and\ \citenamefont {Gelbart}(2009)}]{knobler2009physical}%
  \BibitemOpen
  \bibfield  {author} {\bibinfo {author} {\bibfnamefont {C.~M.}\ \bibnamefont {Knobler}}\ and\ \bibinfo {author} {\bibfnamefont {W.~M.}\ \bibnamefont {Gelbart}},\ }\bibfield  {title} {\bibinfo {title} {Physical chemistry of {DNA} viruses},\ }\href@noop {} {\bibfield  {journal} {\bibinfo  {journal} {Annual review of physical chemistry}\ }\textbf {\bibinfo {volume} {60}},\ \bibinfo {pages} {367} (\bibinfo {year} {2009})}\BibitemShut {NoStop}%
\bibitem [{\citenamefont {Lo{\v{s}}dorfer~Bo{\v{z}}i{\v{c}}}\ \emph {et~al.}(2013)\citenamefont {Lo{\v{s}}dorfer~Bo{\v{z}}i{\v{c}}}, \citenamefont {{\v{S}}iber},\ and\ \citenamefont {Podgornik}}]{lovsdorfer2013statistical}%
  \BibitemOpen
  \bibfield  {author} {\bibinfo {author} {\bibfnamefont {A.}~\bibnamefont {Lo{\v{s}}dorfer~Bo{\v{z}}i{\v{c}}}}, \bibinfo {author} {\bibfnamefont {A.}~\bibnamefont {{\v{S}}iber}},\ and\ \bibinfo {author} {\bibfnamefont {R.}~\bibnamefont {Podgornik}},\ }\bibfield  {title} {\bibinfo {title} {Statistical analysis of sizes and shapes of virus capsids and their resulting elastic properties},\ }\href@noop {} {\bibfield  {journal} {\bibinfo  {journal} {Journal of biological physics}\ }\textbf {\bibinfo {volume} {39}},\ \bibinfo {pages} {215} (\bibinfo {year} {2013})}\BibitemShut {NoStop}%
\bibitem [{\citenamefont {Lo{\v{s}}dorfer~Bo{\v{z}}i{\v{c}}}\ and\ \citenamefont {Podgornik}(2018)}]{lovsdorfer2018anomalous}%
  \BibitemOpen
  \bibfield  {author} {\bibinfo {author} {\bibfnamefont {A.}~\bibnamefont {Lo{\v{s}}dorfer~Bo{\v{z}}i{\v{c}}}}\ and\ \bibinfo {author} {\bibfnamefont {R.}~\bibnamefont {Podgornik}},\ }\bibfield  {title} {\bibinfo {title} {Anomalous multipole expansion: Charge regulation of patchy inhomogeneously charged spherical particles},\ }\href@noop {} {\bibfield  {journal} {\bibinfo  {journal} {The Journal of chemical physics}\ }\textbf {\bibinfo {volume} {149}} (\bibinfo {year} {2018})}\BibitemShut {NoStop}%
\bibitem [{\citenamefont {Bo{\v{z}}i{\v{c}}}\ \emph {et~al.}(2018)\citenamefont {Bo{\v{z}}i{\v{c}}}, \citenamefont {Micheletti}, \citenamefont {Podgornik},\ and\ \citenamefont {Tubiana}}]{bovzivc2018compactness}%
  \BibitemOpen
  \bibfield  {author} {\bibinfo {author} {\bibfnamefont {A.~L.}\ \bibnamefont {Bo{\v{z}}i{\v{c}}}}, \bibinfo {author} {\bibfnamefont {C.}~\bibnamefont {Micheletti}}, \bibinfo {author} {\bibfnamefont {R.}~\bibnamefont {Podgornik}},\ and\ \bibinfo {author} {\bibfnamefont {L.}~\bibnamefont {Tubiana}},\ }\bibfield  {title} {\bibinfo {title} {Compactness of viral genomes: effect of disperse and localized random mutations},\ }\href@noop {} {\bibfield  {journal} {\bibinfo  {journal} {Journal of Physics: Condensed Matter}\ }\textbf {\bibinfo {volume} {30}},\ \bibinfo {pages} {084006} (\bibinfo {year} {2018})}\BibitemShut {NoStop}%
\bibitem [{\citenamefont {Belyi}\ and\ \citenamefont {Muthukumar}(2006)}]{belyi2006electrostatic}%
  \BibitemOpen
  \bibfield  {author} {\bibinfo {author} {\bibfnamefont {V.~A.}\ \bibnamefont {Belyi}}\ and\ \bibinfo {author} {\bibfnamefont {M.}~\bibnamefont {Muthukumar}},\ }\bibfield  {title} {\bibinfo {title} {Electrostatic origin of the genome packing in viruses},\ }\href@noop {} {\bibfield  {journal} {\bibinfo  {journal} {Proceedings of the National Academy of Sciences}\ }\textbf {\bibinfo {volume} {103}},\ \bibinfo {pages} {17174} (\bibinfo {year} {2006})}\BibitemShut {NoStop}%
\bibitem [{\citenamefont {Lo{\v{s}}dorfer~Bo{\v{z}}i{\v{c}}}\ \emph {et~al.}(2012)\citenamefont {Lo{\v{s}}dorfer~Bo{\v{z}}i{\v{c}}}, \citenamefont {{\v{S}}iber},\ and\ \citenamefont {Podgornik}}]{lovsdorfer2012simple}%
  \BibitemOpen
  \bibfield  {author} {\bibinfo {author} {\bibfnamefont {A.}~\bibnamefont {Lo{\v{s}}dorfer~Bo{\v{z}}i{\v{c}}}}, \bibinfo {author} {\bibfnamefont {A.}~\bibnamefont {{\v{S}}iber}},\ and\ \bibinfo {author} {\bibfnamefont {R.}~\bibnamefont {Podgornik}},\ }\bibfield  {title} {\bibinfo {title} {How simple can a model of an empty viral capsid be? charge distributions in viral capsids},\ }\href@noop {} {\bibfield  {journal} {\bibinfo  {journal} {Journal of biological physics}\ }\textbf {\bibinfo {volume} {38}},\ \bibinfo {pages} {657} (\bibinfo {year} {2012})}\BibitemShut {NoStop}%
\bibitem [{\citenamefont {Sivanandam}\ \emph {et~al.}(2016)\citenamefont {Sivanandam}, \citenamefont {Mathews}, \citenamefont {Garmann}, \citenamefont {Erdemci-Tandogan}, \citenamefont {Zandi},\ and\ \citenamefont {Rao}}]{sivanandam2016functional}%
  \BibitemOpen
  \bibfield  {author} {\bibinfo {author} {\bibfnamefont {V.}~\bibnamefont {Sivanandam}}, \bibinfo {author} {\bibfnamefont {D.}~\bibnamefont {Mathews}}, \bibinfo {author} {\bibfnamefont {R.}~\bibnamefont {Garmann}}, \bibinfo {author} {\bibfnamefont {G.}~\bibnamefont {Erdemci-Tandogan}}, \bibinfo {author} {\bibfnamefont {R.}~\bibnamefont {Zandi}},\ and\ \bibinfo {author} {\bibfnamefont {A.}~\bibnamefont {Rao}},\ }\bibfield  {title} {\bibinfo {title} {Functional analysis of the {N}-terminal basic motif of a eukaryotic satellite {RNA} virus capsid protein in replication and packaging},\ }\href@noop {} {\bibfield  {journal} {\bibinfo  {journal} {Scientific reports}\ }\textbf {\bibinfo {volume} {6}},\ \bibinfo {pages} {26328} (\bibinfo {year} {2016})}\BibitemShut {NoStop}%
\bibitem [{\citenamefont {van~der Schoot}\ and\ \citenamefont {Bruinsma}(2005)}]{van2005electrostatics}%
  \BibitemOpen
  \bibfield  {author} {\bibinfo {author} {\bibfnamefont {P.}~\bibnamefont {van~der Schoot}}\ and\ \bibinfo {author} {\bibfnamefont {R.}~\bibnamefont {Bruinsma}},\ }\bibfield  {title} {\bibinfo {title} {Electrostatics and the assembly of an {RNA} virus},\ }\href@noop {} {\bibfield  {journal} {\bibinfo  {journal} {Physical Review E}\ }\textbf {\bibinfo {volume} {71}},\ \bibinfo {pages} {061928} (\bibinfo {year} {2005})}\BibitemShut {NoStop}%
\bibitem [{\citenamefont {Erdemci-Tandogan}\ \emph {et~al.}(2016{\natexlab{b}})\citenamefont {Erdemci-Tandogan}, \citenamefont {Wagner}, \citenamefont {van~der Schoot},\ and\ \citenamefont {Zandi}}]{erdemci2016role}%
  \BibitemOpen
  \bibfield  {author} {\bibinfo {author} {\bibfnamefont {G.}~\bibnamefont {Erdemci-Tandogan}}, \bibinfo {author} {\bibfnamefont {J.}~\bibnamefont {Wagner}}, \bibinfo {author} {\bibfnamefont {P.}~\bibnamefont {van~der Schoot}},\ and\ \bibinfo {author} {\bibfnamefont {R.}~\bibnamefont {Zandi}},\ }\bibfield  {title} {\bibinfo {title} {Role of genome in the formation of conical retroviral shells},\ }\href@noop {} {\bibfield  {journal} {\bibinfo  {journal} {The Journal of Physical Chemistry B}\ }\textbf {\bibinfo {volume} {120}},\ \bibinfo {pages} {6298} (\bibinfo {year} {2016}{\natexlab{b}})}\BibitemShut {NoStop}%
\bibitem [{\citenamefont {Maassen}\ \emph {et~al.}(2018)\citenamefont {Maassen}, \citenamefont {van~der Schoot},\ and\ \citenamefont {Cornelissen}}]{maassen2018experimental}%
  \BibitemOpen
  \bibfield  {author} {\bibinfo {author} {\bibfnamefont {S.~J.}\ \bibnamefont {Maassen}}, \bibinfo {author} {\bibfnamefont {P.}~\bibnamefont {van~der Schoot}},\ and\ \bibinfo {author} {\bibfnamefont {J.~J.}\ \bibnamefont {Cornelissen}},\ }\bibfield  {title} {\bibinfo {title} {Experimental and theoretical determination of the ph inside the confinement of a virus-like particle},\ }\href@noop {} {\bibfield  {journal} {\bibinfo  {journal} {Small}\ }\textbf {\bibinfo {volume} {14}},\ \bibinfo {pages} {1802081} (\bibinfo {year} {2018})}\BibitemShut {NoStop}%
\bibitem [{\citenamefont {Moghaddam}\ and\ \citenamefont {Thormann}(2019)}]{moghaddam2019hofmeister}%
  \BibitemOpen
  \bibfield  {author} {\bibinfo {author} {\bibfnamefont {S.~Z.}\ \bibnamefont {Moghaddam}}\ and\ \bibinfo {author} {\bibfnamefont {E.}~\bibnamefont {Thormann}},\ }\bibfield  {title} {\bibinfo {title} {The hofmeister series: Specific ion effects in aqueous polymer solutions},\ }\href@noop {} {\bibfield  {journal} {\bibinfo  {journal} {Journal of colloid and interface science}\ }\textbf {\bibinfo {volume} {555}},\ \bibinfo {pages} {615} (\bibinfo {year} {2019})}\BibitemShut {NoStop}%
\bibitem [{\citenamefont {Cacace}\ \emph {et~al.}(1997)\citenamefont {Cacace}, \citenamefont {Landau},\ and\ \citenamefont {Ramsden}}]{cacace1997hofmeister}%
  \BibitemOpen
  \bibfield  {author} {\bibinfo {author} {\bibfnamefont {M.}~\bibnamefont {Cacace}}, \bibinfo {author} {\bibfnamefont {E.}~\bibnamefont {Landau}},\ and\ \bibinfo {author} {\bibfnamefont {J.}~\bibnamefont {Ramsden}},\ }\bibfield  {title} {\bibinfo {title} {The hofmeister series: salt and solvent effects on interfacial phenomena},\ }\href@noop {} {\bibfield  {journal} {\bibinfo  {journal} {Quarterly reviews of biophysics}\ }\textbf {\bibinfo {volume} {30}},\ \bibinfo {pages} {241} (\bibinfo {year} {1997})}\BibitemShut {NoStop}%
\bibitem [{\citenamefont {Panahandeh}\ \emph {et~al.}(2018)\citenamefont {Panahandeh}, \citenamefont {Li},\ and\ \citenamefont {Zandi}}]{panahandeh2018equilibrium}%
  \BibitemOpen
  \bibfield  {author} {\bibinfo {author} {\bibfnamefont {S.}~\bibnamefont {Panahandeh}}, \bibinfo {author} {\bibfnamefont {S.}~\bibnamefont {Li}},\ and\ \bibinfo {author} {\bibfnamefont {R.}~\bibnamefont {Zandi}},\ }\bibfield  {title} {\bibinfo {title} {The equilibrium structure of self-assembled protein nano-cages},\ }\href@noop {} {\bibfield  {journal} {\bibinfo  {journal} {Nanoscale}\ }\textbf {\bibinfo {volume} {10}},\ \bibinfo {pages} {22802} (\bibinfo {year} {2018})}\BibitemShut {NoStop}%
\bibitem [{\citenamefont {Yoffe}\ \emph {et~al.}(2008)\citenamefont {Yoffe}, \citenamefont {Prinsen}, \citenamefont {Gopal}, \citenamefont {Knobler}, \citenamefont {Gelbart},\ and\ \citenamefont {Ben-Shaul}}]{yoffe2008predicting}%
  \BibitemOpen
  \bibfield  {author} {\bibinfo {author} {\bibfnamefont {A.~M.}\ \bibnamefont {Yoffe}}, \bibinfo {author} {\bibfnamefont {P.}~\bibnamefont {Prinsen}}, \bibinfo {author} {\bibfnamefont {A.}~\bibnamefont {Gopal}}, \bibinfo {author} {\bibfnamefont {C.~M.}\ \bibnamefont {Knobler}}, \bibinfo {author} {\bibfnamefont {W.~M.}\ \bibnamefont {Gelbart}},\ and\ \bibinfo {author} {\bibfnamefont {A.}~\bibnamefont {Ben-Shaul}},\ }\bibfield  {title} {\bibinfo {title} {Predicting the sizes of large {RNA} molecules},\ }\href@noop {} {\bibfield  {journal} {\bibinfo  {journal} {Proceedings of the National Academy of Sciences}\ }\textbf {\bibinfo {volume} {105}},\ \bibinfo {pages} {16153} (\bibinfo {year} {2008})}\BibitemShut {NoStop}%
\bibitem [{\citenamefont {Chen}\ \emph {et~al.}(2012)\citenamefont {Chen}, \citenamefont {Meisburger}, \citenamefont {Pabit}, \citenamefont {Sutton}, \citenamefont {Webb},\ and\ \citenamefont {Pollack}}]{chen2012ionic}%
  \BibitemOpen
  \bibfield  {author} {\bibinfo {author} {\bibfnamefont {H.}~\bibnamefont {Chen}}, \bibinfo {author} {\bibfnamefont {S.~P.}\ \bibnamefont {Meisburger}}, \bibinfo {author} {\bibfnamefont {S.~A.}\ \bibnamefont {Pabit}}, \bibinfo {author} {\bibfnamefont {J.~L.}\ \bibnamefont {Sutton}}, \bibinfo {author} {\bibfnamefont {W.~W.}\ \bibnamefont {Webb}},\ and\ \bibinfo {author} {\bibfnamefont {L.}~\bibnamefont {Pollack}},\ }\bibfield  {title} {\bibinfo {title} {Ionic strength-dependent persistence lengths of single-stranded {RNA} and {DNA}},\ }\href@noop {} {\bibfield  {journal} {\bibinfo  {journal} {Proceedings of the National Academy of Sciences}\ }\textbf {\bibinfo {volume} {109}},\ \bibinfo {pages} {799} (\bibinfo {year} {2012})}\BibitemShut {NoStop}%
\bibitem [{\citenamefont {Smith}\ \emph {et~al.}(2000)\citenamefont {Smith}, \citenamefont {Chase}, \citenamefont {Schmidt},\ and\ \citenamefont {Perry}}]{smith2000structure}%
  \BibitemOpen
  \bibfield  {author} {\bibinfo {author} {\bibfnamefont {T.~J.}\ \bibnamefont {Smith}}, \bibinfo {author} {\bibfnamefont {E.}~\bibnamefont {Chase}}, \bibinfo {author} {\bibfnamefont {T.}~\bibnamefont {Schmidt}},\ and\ \bibinfo {author} {\bibfnamefont {K.~L.}\ \bibnamefont {Perry}},\ }\bibfield  {title} {\bibinfo {title} {The structure of cucumber mosaic virus and comparison to cowpea chlorotic mottle virus},\ }\href@noop {} {\bibfield  {journal} {\bibinfo  {journal} {Journal of virology}\ }\textbf {\bibinfo {volume} {74}},\ \bibinfo {pages} {7578} (\bibinfo {year} {2000})}\BibitemShut {NoStop}%
\bibitem [{\citenamefont {Silva}\ and\ \citenamefont {Rossmann}(1985)}]{silva1985refinement}%
  \BibitemOpen
  \bibfield  {author} {\bibinfo {author} {\bibfnamefont {A.}~\bibnamefont {Silva}}\ and\ \bibinfo {author} {\bibfnamefont {M.}~\bibnamefont {Rossmann}},\ }\bibfield  {title} {\bibinfo {title} {The refinement of southern bean mosaic virus in reciprocal space},\ }\href@noop {} {\bibfield  {journal} {\bibinfo  {journal} {Acta Crystallographica Section B: Structural Science}\ }\textbf {\bibinfo {volume} {41}},\ \bibinfo {pages} {147} (\bibinfo {year} {1985})}\BibitemShut {NoStop}%
\bibitem [{\citenamefont {Choi}\ \emph {et~al.}(1997)\citenamefont {Choi}, \citenamefont {Lu}, \citenamefont {Lee}, \citenamefont {Wengler},\ and\ \citenamefont {Rossmann}}]{choi1997structure}%
  \BibitemOpen
  \bibfield  {author} {\bibinfo {author} {\bibfnamefont {H.-K.}\ \bibnamefont {Choi}}, \bibinfo {author} {\bibfnamefont {G.}~\bibnamefont {Lu}}, \bibinfo {author} {\bibfnamefont {S.}~\bibnamefont {Lee}}, \bibinfo {author} {\bibfnamefont {G.}~\bibnamefont {Wengler}},\ and\ \bibinfo {author} {\bibfnamefont {M.~G.}\ \bibnamefont {Rossmann}},\ }\bibfield  {title} {\bibinfo {title} {Structure of semliki forest virus core protein},\ }\href@noop {} {\bibfield  {journal} {\bibinfo  {journal} {Proteins: Structure, Function, and Bioinformatics}\ }\textbf {\bibinfo {volume} {27}},\ \bibinfo {pages} {345} (\bibinfo {year} {1997})}\BibitemShut {NoStop}%
\bibitem [{\citenamefont {Fisher}\ and\ \citenamefont {Johnson}(1993)}]{fisher1993ordered}%
  \BibitemOpen
  \bibfield  {author} {\bibinfo {author} {\bibfnamefont {A.~J.}\ \bibnamefont {Fisher}}\ and\ \bibinfo {author} {\bibfnamefont {J.~E.}\ \bibnamefont {Johnson}},\ }\bibfield  {title} {\bibinfo {title} {Ordered duplex {RNA} controls capsid architecture in an icosahedral animal virus},\ }\href@noop {} {\bibfield  {journal} {\bibinfo  {journal} {Nature}\ }\textbf {\bibinfo {volume} {361}},\ \bibinfo {pages} {176} (\bibinfo {year} {1993})}\BibitemShut {NoStop}%
\bibitem [{\citenamefont {Tong}\ \emph {et~al.}(1993)\citenamefont {Tong}, \citenamefont {Wengler},\ and\ \citenamefont {Rossmann}}]{tong1993refined}%
  \BibitemOpen
  \bibfield  {author} {\bibinfo {author} {\bibfnamefont {L.}~\bibnamefont {Tong}}, \bibinfo {author} {\bibfnamefont {G.}~\bibnamefont {Wengler}},\ and\ \bibinfo {author} {\bibfnamefont {M.~G.}\ \bibnamefont {Rossmann}},\ }\bibfield  {title} {\bibinfo {title} {Refined structure of sindbis virus core protein and comparison with other chymotrypsin-like serine proteinase structures},\ }\href@noop {} {\bibfield  {journal} {\bibinfo  {journal} {Journal of molecular biology}\ }\textbf {\bibinfo {volume} {230}},\ \bibinfo {pages} {228} (\bibinfo {year} {1993})}\BibitemShut {NoStop}%
\bibitem [{\citenamefont {Speir}\ \emph {et~al.}(1995)\citenamefont {Speir}, \citenamefont {Munshi}, \citenamefont {Wang}, \citenamefont {Baker},\ and\ \citenamefont {Johnson}}]{speir1995structures}%
  \BibitemOpen
  \bibfield  {author} {\bibinfo {author} {\bibfnamefont {J.~A.}\ \bibnamefont {Speir}}, \bibinfo {author} {\bibfnamefont {S.}~\bibnamefont {Munshi}}, \bibinfo {author} {\bibfnamefont {G.}~\bibnamefont {Wang}}, \bibinfo {author} {\bibfnamefont {T.~S.}\ \bibnamefont {Baker}},\ and\ \bibinfo {author} {\bibfnamefont {J.~E.}\ \bibnamefont {Johnson}},\ }\bibfield  {title} {\bibinfo {title} {Structures of the native and swollen forms of cowpea chlorotic mottle virus determined by {X-ray} crystallography and cryo-electron microscopy},\ }\href@noop {} {\bibfield  {journal} {\bibinfo  {journal} {Structure}\ }\textbf {\bibinfo {volume} {3}},\ \bibinfo {pages} {63} (\bibinfo {year} {1995})}\BibitemShut {NoStop}%
\bibitem [{\citenamefont {Willy}\ \emph {et~al.}(2021)\citenamefont {Willy}, \citenamefont {Ferguson}, \citenamefont {Akatay}, \citenamefont {Huber}, \citenamefont {Djakbarova}, \citenamefont {Silahli}, \citenamefont {Cakez}, \citenamefont {Hasan}, \citenamefont {Chang}, \citenamefont {Travesset}, \citenamefont {Li}, \citenamefont {Zandi}, \citenamefont {Li}, \citenamefont {Betzig}, \citenamefont {Cocucci},\ and\ \citenamefont {Kural}}]{Willy2022}%
  \BibitemOpen
  \bibfield  {author} {\bibinfo {author} {\bibfnamefont {N.~M.}\ \bibnamefont {Willy}}, \bibinfo {author} {\bibfnamefont {J.~P.}\ \bibnamefont {Ferguson}}, \bibinfo {author} {\bibfnamefont {A.}~\bibnamefont {Akatay}}, \bibinfo {author} {\bibfnamefont {S.}~\bibnamefont {Huber}}, \bibinfo {author} {\bibfnamefont {U.}~\bibnamefont {Djakbarova}}, \bibinfo {author} {\bibfnamefont {S.}~\bibnamefont {Silahli}}, \bibinfo {author} {\bibfnamefont {C.}~\bibnamefont {Cakez}}, \bibinfo {author} {\bibfnamefont {F.}~\bibnamefont {Hasan}}, \bibinfo {author} {\bibfnamefont {H.~C.}\ \bibnamefont {Chang}}, \bibinfo {author} {\bibfnamefont {A.}~\bibnamefont {Travesset}}, \bibinfo {author} {\bibfnamefont {S.}~\bibnamefont {Li}}, \bibinfo {author} {\bibfnamefont {R.}~\bibnamefont {Zandi}}, \bibinfo {author} {\bibfnamefont {D.}~\bibnamefont {Li}}, \bibinfo {author} {\bibfnamefont {E.}~\bibnamefont {Betzig}}, \bibinfo {author} {\bibfnamefont {E.}~\bibnamefont {Cocucci}},\ and\ \bibinfo {author} {\bibfnamefont {C.}~\bibnamefont
  {Kural}},\ }\bibfield  {title} {\bibinfo {title} {{De novo endocytic clathrin coats develop curvature at early stages of their formation}},\ }\href {https://doi.org/10.1016/j.devcel.2021.10.019} {\bibfield  {journal} {\bibinfo  {journal} {Developmental Cell}\ }\textbf {\bibinfo {volume} {56}},\ \bibinfo {pages} {3146} (\bibinfo {year} {2021})}\BibitemShut {NoStop}%
\bibitem [{\citenamefont {Timmermans}\ \emph {et~al.}(2022)\citenamefont {Timmermans}, \citenamefont {Ramezani}, \citenamefont {Montalvo}, \citenamefont {Nguyen}, \citenamefont {van~der Schoot}, \citenamefont {van Hest},\ and\ \citenamefont {Zandi}}]{Timmermans}%
  \BibitemOpen
  \bibfield  {author} {\bibinfo {author} {\bibfnamefont {S.~B. P.~E.}\ \bibnamefont {Timmermans}}, \bibinfo {author} {\bibfnamefont {A.}~\bibnamefont {Ramezani}}, \bibinfo {author} {\bibfnamefont {T.}~\bibnamefont {Montalvo}}, \bibinfo {author} {\bibfnamefont {M.}~\bibnamefont {Nguyen}}, \bibinfo {author} {\bibfnamefont {P.}~\bibnamefont {van~der Schoot}}, \bibinfo {author} {\bibfnamefont {J.~C.~M.}\ \bibnamefont {van Hest}},\ and\ \bibinfo {author} {\bibfnamefont {R.}~\bibnamefont {Zandi}},\ }\bibfield  {title} {\bibinfo {title} {The dynamics of viruslike capsid assembly and disassembly},\ }\href {https://doi.org/10.1021/jacs.2c04074} {\bibfield  {journal} {\bibinfo  {journal} {Journal of the American Chemical Society}\ }\textbf {\bibinfo {volume} {144}},\ \bibinfo {pages} {12608} (\bibinfo {year} {2022})},\ \bibinfo {note} {pMID: 35792573},\ \Eprint {https://arxiv.org/abs/https://doi.org/10.1021/jacs.2c04074} {https://doi.org/10.1021/jacs.2c04074} \BibitemShut {NoStop}%
\bibitem [{\citenamefont {Anderson}\ \emph {et~al.}(2020)\citenamefont {Anderson}, \citenamefont {Glaser},\ and\ \citenamefont {Glotzer}}]{anderson2020hoomd}%
  \BibitemOpen
  \bibfield  {author} {\bibinfo {author} {\bibfnamefont {J.~A.}\ \bibnamefont {Anderson}}, \bibinfo {author} {\bibfnamefont {J.}~\bibnamefont {Glaser}},\ and\ \bibinfo {author} {\bibfnamefont {S.~C.}\ \bibnamefont {Glotzer}},\ }\bibfield  {title} {\bibinfo {title} {Hoomd-blue: A python package for high-performance molecular dynamics and hard particle monte carlo simulations},\ }\href@noop {} {\bibfield  {journal} {\bibinfo  {journal} {Computational Materials Science}\ }\textbf {\bibinfo {volume} {173}},\ \bibinfo {pages} {109363} (\bibinfo {year} {2020})}\BibitemShut {NoStop}%
\bibitem [{\citenamefont {Stukowski}(2009)}]{stukowski2009visualization}%
  \BibitemOpen
  \bibfield  {author} {\bibinfo {author} {\bibfnamefont {A.}~\bibnamefont {Stukowski}},\ }\bibfield  {title} {\bibinfo {title} {Visualization and analysis of atomistic simulation data with ovito--the open visualization tool},\ }\href@noop {} {\bibfield  {journal} {\bibinfo  {journal} {Modelling and simulation in materials science and engineering}\ }\textbf {\bibinfo {volume} {18}},\ \bibinfo {pages} {015012} (\bibinfo {year} {2009})}\BibitemShut {NoStop}%
\bibitem [{Note1()}]{Note1}%
  \BibitemOpen
  \bibinfo {note} {The model of ~\cite {van2013impact} maps onto the present model by $\varepsilon \rightarrow - \nu _\pm / l^3$ where $\nu _\pm <0$ is the cross virial coefficient between an N-terminal segment and a chain segment and $l$ denotes the bond length.}\BibitemShut {Stop}%
\bibitem [{\citenamefont {Piculell}\ \emph {et~al.}(1995)\citenamefont {Piculell}, \citenamefont {Viebke},\ and\ \citenamefont {Linse}}]{Piculell1995}%
  \BibitemOpen
  \bibfield  {author} {\bibinfo {author} {\bibfnamefont {L.}~\bibnamefont {Piculell}}, \bibinfo {author} {\bibfnamefont {C.}~\bibnamefont {Viebke}},\ and\ \bibinfo {author} {\bibfnamefont {P.}~\bibnamefont {Linse}},\ }\bibfield  {title} {\bibinfo {title} {Adsorption of a flexible polymer onto a rigid rod. a model study},\ }\href@noop {} {\bibfield  {journal} {\bibinfo  {journal} {Journal of physical chemistry}\ }\textbf {\bibinfo {volume} {699}},\ \bibinfo {pages} {17423} (\bibinfo {year} {1995})}\BibitemShut {NoStop}%
\bibitem [{\citenamefont {Rubinstein}\ and\ \citenamefont {Colby}(2003)}]{rubinstein2003polymer}%
  \BibitemOpen
  \bibfield  {author} {\bibinfo {author} {\bibfnamefont {M.}~\bibnamefont {Rubinstein}}\ and\ \bibinfo {author} {\bibfnamefont {R.~H.}\ \bibnamefont {Colby}},\ }\href@noop {} {\emph {\bibinfo {title} {Polymer physics}}}\ (\bibinfo  {publisher} {Oxford University Press},\ \bibinfo {year} {2003})\BibitemShut {NoStop}%
\bibitem [{\citenamefont {Huang}\ \emph {et~al.}(2021)\citenamefont {Huang}, \citenamefont {Podgornik},\ and\ \citenamefont {Man}}]{Huang2021}%
  \BibitemOpen
  \bibfield  {author} {\bibinfo {author} {\bibfnamefont {C.}~\bibnamefont {Huang}}, \bibinfo {author} {\bibfnamefont {R.}~\bibnamefont {Podgornik}},\ and\ \bibinfo {author} {\bibfnamefont {X.}~\bibnamefont {Man}},\ }\bibfield  {title} {\bibinfo {title} {{Selective Adsorption of Confined Polymers: Self-Consistent Field Theory Studies}},\ }\href {https://doi.org/10.1021/acs.macromol.1c01785} {\bibfield  {journal} {\bibinfo  {journal} {Macromolecules}\ }\textbf {\bibinfo {volume} {54}},\ \bibinfo {pages} {9602} (\bibinfo {year} {2021})}\BibitemShut {NoStop}%
\bibitem [{\citenamefont {Lo~Nostro}\ and\ \citenamefont {Ninham}(2012)}]{Ninham2012}%
  \BibitemOpen
  \bibfield  {author} {\bibinfo {author} {\bibfnamefont {P.}~\bibnamefont {Lo~Nostro}}\ and\ \bibinfo {author} {\bibfnamefont {B.~W.}\ \bibnamefont {Ninham}},\ }\bibfield  {title} {\bibinfo {title} {Hofmeister phenomena: An update on ion specificity in biology},\ }\href@noop {} {\bibfield  {journal} {\bibinfo  {journal} {Chemical Reviews}\ }\textbf {\bibinfo {volume} {112}},\ \bibinfo {pages} {2286} (\bibinfo {year} {202012})}\BibitemShut {NoStop}%
\end{thebibliography}%
\end{document}